\documentstyle[aps,prd,12pt,tighten,amssymb,amsmath,eqsecnum,cite]{revtex}
\def\be{\begin{equation}}
\def\ee{\end{equation}}
\def\bea{\begin{eqnarray}}
\def\eea{\end{eqnarray}}
\def\tr{{\rm tr}}
\def\str{{\rm str}}
\def\parmu{\partial_\mu}

\def\parnu{\partial_\nu}
\def\parrho{\partial_\rho}
\def\ginv{g^{-1}}

\newcommand{\p}{\partial}
\title{\vspace*{1cm}{\large{\bf Towards a Field Theory of the Plateau Transitions in the Integer Quantum Hall Effect}}}
\vspace{0.5cm}
\author{Miraculous J. Bhaseen$^{a}$\footnote{\vspace*{-0.4cm}bhaseen@thphys.ox.ac.uk},
Ian I. Kogan$^{a}$\footnote{\vspace*{-0.4cm}kogan@thphys.ox.ac.uk}, Oleg
A. Soloviev$^{b}$\footnote{\vspace*{-0.4cm}a.soloviev@qmw.ac.uk},\\
Nobuhiko Taniguchi$^{c}$\footnote{\vspace*{-0.4cm}taniguch@cm.ph.tsukuba.ac.jp} and
Alexei M. Tsvelik$^{a}$\footnote{\vspace*{-0.4cm}tsvelik@thphys.ox.ac.uk}
\\ \vspace{0.5cm}$^{a}${\small {\em Theoretical Physics (University of Oxford), \\ 1 Keble Road,
Oxford, OX1 3NP, U.K.}} \\ $^{b}${\small{\em Physics Department, Queen Mary and
Westfield College,\\ Mile End Road, London, E1 4NS,
U.K.}}\\$^c${\small{\em Institute of Physics, University of Tsukuba, Tennodai, Tsukuba 305, Japan.}}}
\begin{document}
\maketitle
\vspace*{-9.5cm}
\begin{flushright}
  OUTP-99-07S\\
QMW-ph-99-20\\
   cond-mat/9912060
\end{flushright}
\vspace*{8cm}

\begin{center}
\today
\end{center}
\vspace{0.5cm}

\begin{abstract}
 We suggest a procedure for calculating
correlation functions of the local densities of states (DOS) 
at the plateau transitions in the Integer
Quantum Hall effect (IQHE). We argue that their  correlation functions  are appropriately described in terms
of the SL($2,{\Bbb C}$)/SU(2) WZNW model (at the usual Ka{\v c}--Moody
point and with the  level $6 \leq k \leq 8$). In this model we have 
identified the operators corresponding to the local DOS, and derived 
the partial differential equation determining their correlation
functions. 
The OPEs for powers 
of the local DOS obtained from this equation are in agreement 
with available results. 
\end{abstract}




\section{Introduction}
\label{sec:intro}
Under the conditions of low temperature and strong perpendicular
magnetic field, a two-dimensional electron gas exhibits a striking
macroscopic manifestation of a quantum phenomenon, namely the quantum Hall effect
\cite{VonKliz:iqhe,Tsui:iqhe}: the Hall conductivity exhibits quantized
plateaus at well defined multiples of $e^2/h$ (a fundamental constant). In samples where the r{\^o}le of
random impurities (disorder) is more important than
electron-electron interactions, the plateaus occur at integer
multiples of $e^2/h$ giving rise to the so-called integer quantum Hall
effect (IQHE). It is widely believed that in the absence of a magnetic
field, all wavefunctions for non-interacting, disordered electrons in
two dimensions are localized. In the presence of a magnetic field
however,  a delocalized state  occurs at the centre of the (disorder
broadened) Landau level, with energy $E_c$. As one tunes the electron
energy $E$ (by varying the magnetic field), through the centre of a
Landau level, the localization length, $\xi$, diverges
as $\xi=|E-E_c|^{-\nu}$, where numerical simulations indicate that $\nu\sim
2.3$.\footnote{We note, that in a system of size $L$, the number of
states with wave functions that reach the boundaries is $\sim
\rho(E)L^{2-1/\nu}$. Since the density of states, $\rho(E)$, remains a
smooth function of energy, this number is always macroscopic. However,
the density of delocalized states $\sim L^{-1/\nu}$ and goes to zero in
the limit of infinite sample size.} The plateaus with differing $\sigma_{xy}$ are  separated  by these critical points.  A theoretical
description of these points remains one of the
most challenging unresolved problems in the theory of disordered systems. 

In the field-theoretic approach to the problem of disordered
electrons, the presence of a static potential preserves the one-body energy, so the frequency 
 becomes simply a parameter of the action  and one may
consider each frequency separately. This reduces the problem of
disordered electrons in $d$ spatial dimensions to a $d$-dimensional
Euclidean field theory. The natural description of the plateau
transitions in the IQHE, should therefore be in
terms of the critical point of some two-dimensional Euclidean field theory, to
which one may apply the powerful machinery of conformal field theory
(CFT) --- see for example the books \cite{Fransesco:cft, Ketov:cft,
Tsvelik:boson}. Despite persistent efforts over the last fifteen
years, the form of the effective action describing the critical point
is still uncertain. 

The first field-theoretical description of the disordered Landau level
was given by Levine, Libby and Pruisken \cite{Levine:sigmod} 
and Pruisken \cite{Pruisken:sigmod} in the form of a non-linear sigma model
with a topological term. On the basis of this model, Khmelnitskii
suggested a two-parameter scaling theory \cite{Khmel:twopar} of the IQHE. In this theory, all the (renormalization group) flow lines
merge into one of a number of fixed points occurring at integer
multiples of the Hall conductance (measured in units of $e^2/h$)
corresponding to the existence of plateaus in the IQHE. In addition,
the flow-diagram contains unstable fixed points, corresponding to the
transition states occurring between plateaus. These flows were 
justified in a dilute
instanton gas approximation by 
 Levine, Libby and Pruisken\cite{Pruisken3} and independently
by Knizhnik and Morozov \cite{Knizmoz}. Unfortunately, the
extrapolation towards the conjectured fixed points lies at strong
coupling, and quantitative results are lacking. In the original 
derivation of the model, Pruisken
and collaborators employed the method of replicas, leading to a sigma
model defined on the manifold U($2n$)$/$U($n$)$\times$U($n$) with
$n\rightarrow0$. A more rigorous formulation using Efetov's
supersymmetry approach \cite{Efetov:super,Efetov:chaos} was given 
in \cite{Weiden:supersig}, and led to a sigma
model on the manifold SU($1,1|2$)/U($1|1$)$\times$ U($1|1$). The action
may be written in the following form, 
\begin{equation}
\label{pruisken}
S=\int d^2 x\, \str\left[-\frac{1}{8\alpha}\left(\partial_\mu Q\right)^2+\frac{1}{8}\sigma^0_{xy}\left(Q\left[\partial_x Q,\partial_y Q\right]\right)+\frac{1}{2}\pi\rho_0\eta\Sigma^3Q\right], 
\end{equation}
where $Q$ is a $4\times4$ supermatrix satisfying the conditions,
\begin{equation}
\quad \str Q=0,\quad Q^2=1,
\end{equation}
$\Sigma^3={\rm diag}(1,-1,1,-1)$ (in the boson-fermion
supermatrix representation), $\sigma^0_{xy}(E)$ is the bare Hall
conductance at energy $E$, $\rho_0$ is the average density of states 
at energy $E$,  and $\eta$ is the imaginary frequency 
which serves as a symmetry
breaking field.   At weak
disorder, $\alpha\ll 1$, the bare value of the inverse coupling constant,
$\alpha^{-1}$, coincides with the  bare longitudinal conductance 
$\sigma^{0}_{xx}(E)$ at energy $E$. To be more precise, one can
directly relate the bare parameters of the action to experimentally
observable quantities of a mesoscopic system of size $L \sim l$ (the
mean free path). At $E=E_c$, one has
$\sigma^0_{xy}=\frac{1}{2}$, and the model is  expected to 
be critical at large
distances.


The second term appearing in (\ref{pruisken}) is topological, and
despite the fact that its presence is crucial for the critical
behaviour, its effect cannot be spotted in a perturbative expansion in
powers of $\alpha$; it does not contribute to the equations of motion and
hence does not contribute to the loop expansion of the beta
function. The effects of the topological term become visible only for
samples of size greater than $\xi\sim l\exp[\pi{\sigma^0_{xx}}^2]$,
where $l$ is the electron mean free
path. In the model (\ref{pruisken}),  with $\sigma^0_{xy}=0$, the length
scale $\xi$ corresponds to the localization length; 
with $\sigma^0_{xy}=1/2$, this scale is the 
{\em transmutation length} (in field-theoretic jargon) and signifies a
crossover to the regime of {\em universal} critical fluctuations.

It is widely believed  that the model (\ref{pruisken}) 
has  a non-trivial
infrared fixed point at some $\alpha^*$ and $\sigma_{xy} =
1/2$. However,  one
cannot  simply substitute $\alpha^*$ with the experimental value of
 $(\sigma_{xx}^\ast)^{-1}$ (the inverse longitudinal conductance) 
in order to obtain the effective action at the critical point.
The reason for this is
that the fixed point occurs at strong coupling, where the fundamental
fields of the model (\ref{pruisken}) are strongly fluctuating. 
In this regime, the  coupling constant $\alpha$, being dependent  on the
regularization procedure,  
can  no longer be identified with any measurable  quantity. 
This phenomenon is well known in
asymptotically free gauge theories --- the gauge coupling
constant has meaning  only at short distances, and cannot be defined in
a universal way in the infrared.\footnote{Notable exceptions to the rule
 are theories with spontaneously broken gauge symmetry (in which case 
zero-charge
  U(1) can also emerge) where the coupling constant can be defined 
in the infrared
through  either an effective  Fermi four-fermion interaction or 
Thompson scattering.} In QCD, for example, gluons do
not exist as asymptotic states in the infrared --- there is  a mass
transmutation phenomenon and the emerging degrees of freedom are
massive. In the quantum Hall effect, however, we expect the emerging infrared theory to be a massless CFT, and so one should not stretch the analogy too far. A better analogy is the Seiberg duality \cite{Seiberg:seiberg} appearing in $N=1$ supersymmetric
 gauge theories, with $N_c$ colours and $N_f$ flavours, in the so-called
 conformal window, $3N_c/2 < N_f < 3N_c$.  There is an infrared (IR)
  fixed point in this theory, in the vicinity of which, the theory 
can be described as an
  infrared limit of {\em another} gauge theory (with the same number 
of flavours and $N_f-N_c$
 colours). If $N_f$ is close to $3N_c/2$ the original theory is 
strongly coupled near the IR fixed
  point, whereas the second is only weakly coupled; the appropriate  IR
behaviour is given by the second (dual) formulation.

Another example of this nature occurs in the deformation of the minimal model $M_p$ of
two-dimensional CFT, by the operator $\Phi_{1,3}$ 
\cite{Zamol, Ludwig}. The theory in the infrared can be described
either as a strongly fluctuating asymptotically free $M_p$ model, or
as a weakly fluctuating infrared free $M_{p-1}$ model, deformed by the
irrelevant operator $\Phi_{3,1}$. Again, it is the second (dual)
description which is required in the infrared.

 These analogies encourage us to believe that the theoretical
description of the plateau transitions in the IQHE will ultimately
lie in a reformulation of the model (\ref{pruisken}), in terms of
degrees of freedom more appropriate for the infrared region. 

An alternative approach to the study of the plateau transitions is
based on the Chalker--Coddington Network Model \cite{Chalker:net}. This
model has been reformulated in Hamiltonian form as a Replica Spin
Chain \cite{DhLee:94,KondMar:97}, and as a Superspin Chain
\cite{Read:super,Zirn:super,Zirn:97,Zirn:99}. The Superspin Chain, is
a lattice model with nearest neighbour antiferromagnetic exchange
between `spins', which are generators of the gl($2|2$) algebra. The advantage of such reformulations is that one may approach
the critical point  more accurately by tuning the parameters  of the
network. This makes the model indispensable for numerical
simulations. However, the Chalker--Coddington model (or the spin chain
models) are lattice theories, and one still needs to derive their
continuum limits in order to describe the critical fluctuations. The
derivation of such limits remains an open problem.

Exasperated by long and futile efforts to find a rigorous derivation
of the effective action at the critical point, Zirnbauer has drawn
upon a great many sources and conjectured its possible form 
\cite{Zirnbauer:integer}\footnote{Here we would like to mention the 
 recent paper \cite{guruswamy} containing  interesting suggestions
with respect to 
derivation of the critical theory.}. Such a 
conjecture ought to be justified, amongst other
considerations, by its ability to reproduce existing results for
physical quantities at the critical point.  
 In \cite{Zirnbauer:integer} this has been successfully  done 
 for the two-point
conductance. In the present  paper we shall concentrate on 
 the correlation functions
of the local density  of states (DOS).

The structure of our paper is as follows: in section
\ref{sect:corrandope} we discuss a possible form for the correlation
functions at the transitions. In particular, we
discuss the operator product expansion (OPE) for moments  of the local
density of states  $\rho^q$, which reflects the multifractality
of the critical wavefunctions. The two-point correlation function
of the local DOS is given as a particular limit of a four-point
function. 
The latter function
includes two  operators with anomalous dimension zero (vacuum insertions)
which serve to redefine the ground state of the theory. In section
\ref{dicorep}
we discuss more fully the important r\^ole played by these operators. In section \ref{sect:onedim} we discuss a possible connection between
the properties of the required {\em two-dimensional} conformal field
theory, and non-perturbative results \cite{Mirlin:onedim} obtained by studying the {\em one-dimensional}
limit of model (\ref{pruisken}). The analysis of the one-dimensional
case brings us to the idea that one can study the behaviour of the local
DOS in terms of the SL($2,{\Bbb C}$)/SU(2) Wess--Zumino--Novikov--Witten
(WZNW) model. In sections \ref{sect:gauged} and \ref{sect:decoup} we study one of the
candidate theories for describing the plateau transitions, namely the
PSL($2|2)$ WZNW model
\cite{Zirnbauer:integer}. The non-compact bosonic sector of this theory is the
SL($2,{\Bbb C}$)/SU(2) WZNW model. In section \ref{sect:dos} we use the SL($2,{\Bbb C}$)/SU(2) WZNW model to
study the correlation functions of the local DOS. Finally we present
concluding remarks (including open problems) and technical appendices.

\section{Constraints on Candidate Theories}
\label{sect:corrandope}
Any conformal field theory purporting to describe the plateau
transitions must satisfy a number of constraints, some of which 
have been discussed by Zirnbauer
\cite{Zirnbauer:integer}. For the purpose of this paper, we believe
the following are important requirements:
\begin{enumerate}
\item{The model should be defined over a manifold associated with 
some supergroup, and must contain a non-trivial (non-unit)
operator with zero scaling dimension (see, for example \cite{Efetov:super}).} 
\item The model must reproduce all known scaling dimensions. 
\item Restricting our two-dimensional CFT to a finite-width strip
geometry (in a suitable limit to be discussed below, and in section
\ref{sect:onedim}) one should recover certain known
quasi-one-dimensional results.  
\end{enumerate}
We shall now take each of these constraints in turn, and consider some
of their consequences.
\subsection{Supermanifold and Zero-Dimension Operator}
\label{supzerodim}
Given our belief that a (presently unknown) non-perturbative treatment of model
(\ref{pruisken}) will ultimately describe the plateau
transitions in the IQHE, it is entirely natural to expect the
resulting effective action to be defined over some
supermanifold. However, in order to combine conformal symmetry with
bosonic and fermionic degrees of freedom, one will ultimately face a
zero-mode problem: if the effective action at the critical point
contains only terms with derivatives, the integral over the zero modes
(constant fields) leaves the partition function
ill-defined. Explicitly,
\begin{equation}
\int dQ_0=\int dB\int dF
\end{equation}
is indeterminate since the integral over the fermionic fields,
$F$, is zero, whereas the integral over the bosonic fields, $B$, is
infinite (assuming that manifold is non-compact as in the original
model (\ref{pruisken})). In the sigma model (\ref{pruisken}) this
problem is solved by the presence of the additional term
\begin{equation}
S_{\eta} = \eta\int d^2x\, \str(\Sigma^3Q) \label{eta}
\end{equation} 
This term is responsible for energy level broadening in the original
theory. The corresponding  operator has zero scaling dimension\footnote{We note that the present theory is a
purely spatial one, and all dimensions are expressed in terms of units
of length, not time. From this point of view time is dimensionless and
the frequency, $\eta\sim1/[{\rm Time}]$, corresponds to a
dimensionless operator.} and
its value does not change in all orders of perturbation theory. The
term is highly relevant
and generates a {\em
length scale} $L_\eta\sim\eta^{-1/2}$. 

 The  zero modes are also eliminated when one considers an open
 system. The  possibility of  escape through the boundaries 
gives rise to non-uniform
 (i.e. energy-dependent) broadening of the energy levels. For this reason
 closed and open systems may require slightly different approaches. 
In this paper we  consider  only {\it closed}
 systems and  shall  not discuss any  problems related to the 
 conductance. In a  closed system one  has 
to introduce boundary conditions as  was
suggested by Zirnbauer \cite{Zirnbauer86}
and subsequently used by 
various authors  \cite{Mirlin:onedim,Falko}. 
In the field theory approach  
these boundary conditions can be
realized by the  insertion of certain operators into correlation
functions. 
Such insertions naturally lead to a dependence of the correlation
functions on  the distance $L_B$ from the boundary. However, if the
boundary operator contains a non-trivial  zero
dimension operator (we shall call this operator $\Psi_0$), one can
obtain 
a finite answer for the correlation function on sending the boundaries to
infinity.

\subsection{Local DOS}
 
 In a  closed system one cannot study conductance. The only remaining
 quantities to study are the wave
 functions. Since wave functions are not observables, 
the original sigma model does not generate them 
 directly, but rather allows one to calculate correlation 
  functions of the local density  of states 
(DOS). 
%
We believe that  these  correlation functions can 
be expressed solely in terms of
the (non-compact) bosonic sector of the full theory. (For the
one-dimensional case this point is discussed in 
\cite{Efetov:super}, and for two dimensions in
\cite{Falko}.) 

 In  a system of finite
size where all energy levels are discrete,  the local DOS is
defined as 
\begin{equation}
\rho(x,E) = \sum_a|\psi_a(x)|^2\delta(E - E_a)
\end{equation}
where $a$ denotes eigenstates of the system. However, 
this definition is not
suitable  for calculation of averages of the local DOS because products
of delta functions yield infinities. These infinities are removed if
one introduces a finite level broadening replacing the delta functions
by the Lorentzians:
\begin{equation}
\delta(E - E_a) \rightarrow \frac{1}{\pi}\frac{\eta}{(E - E_a)^2 +
\eta^2}
\end{equation}
As we have mentioned above, in the sigma model the level broadening 
comes from term (\ref{eta}). This term does not break  conformal
invariance provided  the level broadening is
much smaller than the mean level spacing, $\eta L^2 << 1$. 
Even such small broadening
does the trick of removing infinities from correlators of the local DOS.   
In what follows we consider {\it normalized} powers of  the DOS defined as 
\begin{equation}
[\rho]^q = (\eta\rho)^q \label{normalized}
\end{equation}
In the context
of the sigma model (\ref{pruisken}) the normalized $[\rho]^q$ is
identified with $\{\eta \str(\Sigma Q k)\}^q$ where $k$ is a  certain
 constant
matrix. 

 In a closed system the wave function
on the boundary does not depend on the fields, that is equal to
unity. However, it may occur (and we shall further 
explain why this is the case for the theories in question) that 
the unit operator is not a primary field
 of the problem. Then one needs to decompose the unit operator 
 in terms of the primaries:
\bea
I(z) = \Psi_0(z) + \int d\mu(p) \Psi_p(z) 
\eea
In this decomposition $\Psi_0$ has zero conformal dimension and the
other fields $\Psi_p(z)$  have positive dimensions \footnote{An analogue of this decomposition 
  is equation (3.7)  of \cite{Mirlin:onedim}:
\begin{eqnarray*}
W^{(1)}(X,\tau) = 2 X^{1/2}\left[K_1(2X^{1/2}) + \frac{2}{\pi}
 \int_0^\infty
 dp \frac{p}{1+p^2} \sinh \frac{\pi p}{2} K_{ip}(2X^{1/2})
\exp(-\frac{1+p^2}{4}\tau)\right]
\end{eqnarray*}
where $K_n$ is the modified Bessel function of the third kind and 
 $W^{(1)}(X,\tau)$ is a solution of a differential equation
 \begin{displaymath}
\left(X^2\frac{\partial^2}{\partial
X^2}-X\right)W^{1}(X, \tau)= 
\frac{\partial W^{(1)}(X, \tau)}{\partial \tau} \label{sch}
\end{displaymath}
with the boundary condition $W^{(1)}(X, 0) = 1$; here $\tau$ plays the
role of $z$. 

We see that at zero $\tau$ this is indeed the decomposition of unity
 and at infinite $\tau$ only the first term $ 2 X^{1/2}K_1(2X^{1/2}) $
survives - which is an analogue of $\Psi_0(z)$.}.

 Working with the normalized DOS
and the boundary operators 
 present one can define the correlation functions in the infinite
system (we imagine this system as a long strip  of length $2L$ with
periodic boundary conditions in the transverse direction) ). In  the limit
\begin{equation}
\lim_{L \rightarrow \infty \atop \eta L^2 \rightarrow
0}\langle I(- L)(\eta\rho)^q(x)(\eta\rho)^q(y)I(L)\rangle = \lim_{L \rightarrow \infty}\langle\Psi_0(-L)[\rho^q](x)[\rho^q](y)\Psi_0(L)\rangle\label{limit}
\end{equation}
only $\Psi_0$ operators
survive as $L \rightarrow \infty$.  
 Taking this into account and also that expression (\ref{limit}) 
is valid in the strip  geometry, one obtains  the following expression for
the two-point correlation function in the infinite  plane: 
\begin{equation} 
\overline{[\rho^q]({\bf r}_1)[\rho^q]({\bf r}_2)}= \lim_{|{\bf r_3}|\rightarrow 0, |{\bf r}_4|\rightarrow\infty}\langle[\rho^q]({\bf
r}_1)[\rho^q]({\bf r}_2)\Psi_0({\bf r}_3)\Psi_0({\bf r}_4)\rangle
\end{equation}
Here  we use the overbar to denote disorder averaging, and use the
angular brackets to denote the correlation functions of our field
theory. In  the case of strip  geometry which can be obtained
from the plane  by a conformal transformation, the point of origin is
mapped onto minus infinity.  We shall return to a more detailed 
discussion of this procedure towards  the end of Section IV. 

\subsection{Scaling Dimensions}
\label{subscaldim}
The theory we seek should reproduce all known scaling dimensions
associated with the plateau transition in the IQHE. 
The most famous of these is the localization length exponent,
$\nu$, mentioned in the introduction.\footnote{If we assume that the correlation
length is generated by a single operator, of scaling dimension $d$
coupled to ($\sigma_{xy}-1/2$) and assume further a linear
relationship between the Hall conductance and $|E-E_c|$ in the vicinity
of the transition, we conclude that the scaling dimension of the
operator is 
$d=2-1/\nu\sim 1.57$.}

Other available information comes from the behaviour of the local
DOS, $\rho(E,{\bf r})$, in a two-dimensional sample of side $L$. At the critical point, $\rho({\bf
r})=\rho(E=E_c,{\bf r})$, they are known to satisfy the following fusion
rules\footnote{We adopt the usual CFT nomenclature in anticipation of
the Virasoro-primary nature we shall subsequently motivate for the
local DOS.}\cite{Wegner}:
\begin{eqnarray}
\label{discfusion}
\rho^{p}({\bf r}_1)\rho^{q}({\bf r}_2) & \sim & |{\bf r}_1-{\bf r}_2|^{-d(p)-d(q)+2d(p+q)}\rho^{p+q}({\bf r}_2)\\[0.3cm]
 \overline{\rho^{p}({\bf r}_1)} & \sim & L^{-d(q)} 
\end{eqnarray}
where, once again, the bar stands for disorder averaging; we reserve angle brackets
for field-theoretic correlation functions. (Recall that the local DOS are 
normalized in such a way that their correlation functions are finite
in the limit of vanishing level broadening.) Numerical simulations (see \cite{Janssen:mulfrac} and references
therein) indicate that\footnote{ The proportionality constant in this
expression is denoted as $2/k$ with some hindsight: the parameter $k$
will ultimately be identified as the level of the SL(2,${\Bbb C}$)/SU(2) WZNW theory.} (see Appendix \ref{multi}),
\begin{equation}
\label{scalrel}
d(q)=2q(1-q)/k, \quad 2/k=0.28\pm 0.03.
\end{equation}
This gives $6 < k <8$.

The quadratic dependence of $d(q)$ gradually becomes linear for $|q|>2.5$. We note that the scaling dimensions become negative for
$q<0$ and $q>1$, well inside the interval of validity
(\ref{scalrel}). This suggests that the conformal field theory we are
looking for is {\em non-unitary}.  
\subsection{Quasi-One-Dimensional Results}
Transfer matrix calculations for model (\ref{pruisken}) in a quasi-one-dimensional,
closed (no external leads), infinite sample, yield the following
 form for the two-point correlation function of the local
DOS (see equations (3.58), (3.63) and (3.64) of \cite{Mirlin:onedim}):
\begin{equation}
\label{corrfun}
\overline{[\rho^q](x_1)[\rho^q](x_2)}=\int\,d\mu(p)\,|\langle 0|
Q^q|p\rangle|^2 \exp\Bigl(-|x_{12}|{\tilde d}(p)/4\xi\Bigr),
\end{equation}
where
\begin{eqnarray}
{\tilde d}(p) & = & 1+p^2 \label{contin}\\
d\mu(p) & = & \frac{2p\sinh(\pi p)}{\pi^2}dp \label{measure}\\
\langle 0| Q^q|p\rangle & = & \int_0^\infty\, dX\, X^{q-2}\,
W_{0}(X)W_{(-1+ip)/2}(X)  \label{insight}\\
&\sim &   \left|\Gamma[q + (1 + ip)/2]\right|^2\left|\Gamma[q - (1 +
ip)/2]\right|^2 \nonumber 
\end{eqnarray}
and $\xi$ is the correlation length. The functions $W_{j}$ satisfy the
following eigenvalue equation
\begin{equation}
\left(X^2\frac{\partial^2}{\partial
X^2}-X\right)W_{j}(X)=j(j+1)W_{j}(X). \label{W213}
\end{equation}
The solutions to this equation may be written in the form
\begin{equation}
\label{bessel}
W_j(X)=2{\sqrt X}K_{(1+2j)}(2\sqrt X)
\end{equation}
where $K_p$ is the modified Bessel function of the third kind. 
The relevance of these
quasi-one-dimensional calculations for the critical theory in two
dimensions, will be discussed in detail in section
\ref{sect:onedim}. We simply note here that the correspondence
emerges when one places the CFT on a strip of width $2\pi R$. If the
infinite plane is parametrized by the coordinates $(z,{\bar z})$,
and the strip by $(w,{\bar w})$, one 
may effect this mapping by means of the conformal transformation
$w=R\ln z$. 
Under certain conditions (in
particular, one has to consider $R\sim\xi$) such a strip may be
regarded as a quasi-one-dimensional wire.    
\section{Discrete and Continuous Representations}
\label{dicorep}
To the reader familiar with conformal field theory, equations
(\ref{discfusion}) and (\ref{corrfun}) may appear
contradictory. If $\rho^q$ is a primary field, under what
representation of the symmetry group of the theory does it transform? Whatever the symmetry group of the `grand' theory turns
out to be, we expect it to be some subgroup (or coset) of
Gl($2|2$), and as such may support both discrete and continuous
representations. Equation (\ref{discfusion}) suggests that $\rho^q$ transforms according to a discrete representation, whereas equation
(\ref{corrfun}) suggests that $\rho^q$ is not a single primary
field, but a linear combination of fields with different scaling
dimensions.\footnote{A
 simple resolution of the paradox that the integral (\ref{corrfun})
 generates  simple powers is denied by the 
 analytical properties of the exponential function which prevent one
 from deforming the contour of integration onto the poles of the matrix
 element (\ref{insight}).}

We suggest a resolution of this paradox which is independent of any
particular form of the plateau theory. The solution is intimately
connected to the zero-mode problem and the existence of a
(non-trivial) zero-dimension operator discussed in section
\ref{supzerodim}. 
We  suggest that the two-point
function of the {\it normalized} (see the 
discussion around Eq.(\ref{normalized}))  local DOS 
is given by the limit of the 
{\em four-point} function
of some CFT, 
\begin{equation} 
\overline{[\rho^q]({\bf r}_1)[\rho^q]({\bf r}_2)}= \lim_{|{\bf r_3}|\rightarrow 0, |{\bf r}_4|\rightarrow\infty}\langle[\rho^q]({\bf
r}_1)[\rho^q]({\bf r}_2)\Psi_0({\bf r}_3)\Psi_0({\bf r}_4)\rangle.
\end{equation}

 Let us discuss this suggestion in detail.  
Invariance under the projective
transformations of the plane\footnote{Namely transformations of the
form $w = (az + b)/(cz + d)$, where $ad - bc = 1$.}, restricts the
above four-point function (which we denote by $G_q(1,2,3,4)$) to have the form
\begin{equation}
\label{four}
G_q(1,2,3,4)= \frac{1}{|z_{12}|^{4h_q}}{\cal F}_q(z,\bar z); 
\quad \quad z = \frac{z_{32}z_{41}}{z_{31}z_{42}},
\end{equation}
where we have used the fact that $\rho^q$ has conformal
dimension $h_q (= d_q/2)$, and $\Psi_0$ has zero conformal dimension. 
By definition, the function (\ref{four}) is invariant under the
permutation of $z_1$ and $z_2$ which implies
\begin{equation}
{\cal F}_q(z, \bar z) = {\cal F}_q(1/z,1/\bar z). \label{crossing}
\end{equation}
 In order to reproduce the fusion rules  of the
local DOS (\ref{discfusion}) one  needs to consider the limit, $z_{12}
\rightarrow 0$. The three-point correlation functions are fixed by conformal
invariance, their holomorphic dependence being given by
\begin{equation}
\langle {\cal O}_{h_1}(1){\cal O}_{h_2}(2){\cal O}_{h_3}(3)\rangle = C_{123}\,z_{12}^{- h_1 - h_2 +
h_3}z_{13}^{- h_1 - h_3 + h_2}z_{23}^{- h_2 - h_3 + h_1}
\end{equation}
where $C_{123}$ are the so-called structure constants of the
theory. Using this fact,  one  obtains the desired limit
\begin{eqnarray}
\langle[\rho^q](1)[\rho^q](2)\Psi_0(3)\Psi_0(4)\rangle & \rightarrow & |z_{12}|^{-4h_q +
2h_{2q}}\,\langle[\rho^{2q}](2)\Psi_0(3)\Psi_0(4)\rangle \nonumber\\
& = &  |z_{12}|^{-4h_q +2h_{2q}} \,C_{2q}^{00}\,|z_2|^{- 2h_{2q}},
\end{eqnarray}
 (recall that $z_3 = 0, z_4 \rightarrow \infty$). This result fixes 
 the asymptotics of ${\cal F}_q(z)$ as $1-z \rightarrow 0$ in the
 following manner:
\begin{equation}
{\cal F}_q(z) \sim (1 - z)^{h_{2q}}.\label{asympt1}
\end{equation}
As we shall  now demonstrate,  
 the two-point function of the local DOS 
 in the strip geometry 
explores  different asymptotics of ${\cal F}_q(z)$. One can  map 
 the plane to the strip by the transformation
$w=R\ln z$; point 3 goes to $ - \infty$ and point 4 goes to $
+\infty$. As a result one obtains   (the holomorphic part of) 
 the correlation function as follows:
\begin{equation}
G_q(1,2,3,4) = [2R\sinh(w_{12}/2R)]^{-2h_q}{\cal F}_q\left[\exp(w_{12}/R)\right].
\end{equation}
Notice that  in the strip geometry the correlation function is
translationally invariant. In the limit   ${\rm Re}\, w_{12}\gg
 R$ the behaviour of this function is governed  by the asymptotics  of
${\cal F}_q(z)$ for  $|z|\gg 1$.  In   the limit ${\rm Re}\, w_{21} \gg
 R$ it is
 governed  by the 
asymptotics for  $|z| \ll 1$. (These two limits are related by the crossing invariance
condition (\ref {crossing}).) We see that the fusion rule (\ref{discfusion}) and   the expansion in continuous
representations (\ref{corrfun}) appear in {\em different  channels} of
the four-point function. 

 Thus the two-point function of the local DOS explores different
 asymptotics of the function ${\cal F}$ in different geometries. In this way one may to expect to resolve  the paradox. 

 We conclude this Section with a list of Operator Product Expansions
(OPE) which are necessary to reproduce the above results. First we shall specify our normalization conventions. We normalize all two-point
correlation functions of primary fields ${\cal O}_h$ from {\em discrete}
representations as follows:
\begin{equation}
\langle {\cal O}_h(1){\cal O}_{h'}(2)\rangle = \delta_{h,h'}\,z_{12}^{-2h}.
\end{equation}
The operators from continuous representations ${\cal V}_p(z)$ 
are normalized by
the invariant measure $d\mu/dp$ on the group\footnote{It is highly
likely that for the true theory, the measure coincides with the one
given by equation (\ref{measure}) and $h(p)  = c(1
+ p^2)/8$. However, in this general discussion we do not need to be so specific.}:
\begin{equation}
\langle{\cal V}_p(1){\cal V}_{-p'}(2)\rangle = \frac{\delta(p -
p')}{d\mu/dp}z_{12}^{- 2h(p)}.
\end{equation}
Adopting these conventions, one may write the OPEs in the following manner:
\begin{eqnarray}
[\rho^q](1)[\rho^p](2) & = & C_{qp}^{(q + p)}\,|z_{12}|^{- 2h_q - 2h_p + 2h_{(q
+ p)}}\,[\rho^{(q + p)}](2) + \cdots
\end{eqnarray}
\begin{eqnarray}
[\rho^q](1)\Psi_0(2) & = & |z_{12}|^{-4h_q}\,C_{00}^q\,\Psi_0(2) + \int d\mu(p) |z_{12}|^{- 4h_q +
2h(p)}C^q_{0,q}(p){\cal V}_{p}(2) + \cdots
\end{eqnarray}
where the ellipsis stand for less singular terms. With these conventions
one finds the following asymptotics for ${\cal F}(z)$:
\begin{eqnarray}
{\cal F}(z \rightarrow 0)& = &C_{qq}^{2q}\,C_{00}^{2q}\,|z|^{2h_{2q}} + \cdots \\
{\cal F}(z \rightarrow \infty) & =& |z|^{4h_q}\left[[C_{00}^{q}]^2 + \int d\mu(p)\,|C_{0,p}^{q}|^2\,|z|^{-2h(p)} + \cdots\right].
\end{eqnarray}

\section{CFT on the Strip}
\label{sect:onedim}
\subsection{General Considerations}
In the absence of a rigorous derivation of the critical model describing
the plateau transitions in the IQHE, one is forced to make some assumptions
about the general form of the action. Given the form of the model
(\ref{pruisken}), it is natural to assume that this action is of the
sigma model type, and probably includes the Wess--Zumino term. The reason for including the Wess--Zumino term is related to
the fact that nearly all critical sigma models that we know of
require such a term to ensure criticality. Models of this kind are known as
 Wess--Zumino--Novikov--Witten (WZNW) models; their actions either have a
full group symmetry, $G$, or are defined on a coset space, $G/H$, with
$H$ being a subgroup of $G$. The symmetry manifold of the required
WZNW model is some supergroup manifold, whose symmetry is almost
certainly greater than that of the original model (\ref{pruisken}).

The WZNW action may be written in the form
\begin{equation}
\label{wznw0}
S=\int d^2x \left[\sqrt g g^{\mu\nu}G_{ab}[X]\partial_\mu
X^a\partial_\nu X^b+\epsilon^{\mu\nu}B_{ab}[X]\partial_\mu
X^a\partial_\nu X^b\right]
\end{equation}
where $X^{a}$ are fields representing the coordinates on some group
(or coset) manifold, $G_{ab}$ is the metric tensor on this
manifold, and $B_{ab}$ is an antisymmetric tensor. We define the action
on a curved (world sheet) surface with a metric
$g_{\mu\nu}$ $({\mu,\nu}=1,2)$.

A very important feature of the action (\ref{wznw0}) is that the
second term does not contain the world sheet metric. Consequently, the
classical stress-energy tensor, $T_{\mu\nu} = \delta S/\delta
g^{\mu\nu}$, is determined solely by the first term. In particular, the most important
components for the critical model are given by,
\begin{eqnarray}
T_{zz} = G_{ab}[X]\p_{z}X^a\p_{z}X^b, \quad  T_{\bar z\bar z} =
G_{ab}[X]\p_{\bar z}X^a\p_{\bar z}X^b\label{stress}
\end{eqnarray}
where $z = x_0 + ix_1, \bar z = x_0 - ix_1$. 
 Here, the reader should not  get the false impression that the Wess--Zumino
term is not important. Unlike the {\em topological term} in action
(\ref{pruisken}) it does contribute to the equations of motion. Since
the model (\ref{wznw0}) is supposed to be {\em 
critical}, these equations (to  be understood as identities for
correlation functions in the quantum theory) are 
\begin{eqnarray}
T_{z\bar z} = 0, ~~ \p_{\bar z}T_{zz} = 0, ~~ \p_z T_{\bar z\bar z} =
  0.
\end{eqnarray}
Their fulfillment depends on the Wess--Zumino term through the 
dynamics of the underlying fields $X^a$. 

The smallness of $1/k$ means that there are many fields in the
 theory with conformal dimensions much smaller than unity. This gives weight to the idea that the critical point occurs in the region
 where the coupling constant of the sigma model is relatively
 small. Thus, one may attempt to describe the critical point using the
 semiclassical approximation. We shall formulate the semiclassical
approximation with the specific aim of establishing contact 
between our calculations, and the
 calculations of correlation functions of the 
local DOS for quasi-one-dimensional
 systems performed by Mirlin \cite{Mirlin:onedim} and
 Fyodorov and Mirlin\cite{MF}.

In the infinite plane, parametrized by coordinates $(z,{\bar z})$, conformal
invariance restricts the two-point function of primary fields of
conformal dimension ($h$,${\bar h})$ to be of the form
\begin{equation} 
\label{twopoint}
\langle\phi(z_1,{\bar z}_1)\phi(z_2,{\bar z }_2)\rangle=(z_1 -
z_2)^{-2h}(\bar z_1 - \bar z_2)^{-2\bar h}.
\end{equation}
Under a conformal transformation of the plane, $w=w(z)$, this correlation function
transforms like a tensor of rank $(h,{\bar h})$:
\begin{equation}
\label{tensortran}
\langle \phi(w_1,{\bar w}_1)\phi(w_2,{\bar w}_2)\rangle 
=\prod_{i=1}^2\left(\frac{d z}{dw}\right)^{h}_{w_i}\left(\frac{d
\bar z}{d\bar w}\right)^{\bar h}_{\bar w_i}\langle\phi(z_1,{\bar z}_1)\phi(z_2,{\bar z }_2)\rangle.
\end{equation}
One may pass from the infinite plane to a strip of width $2\pi R$ by
means of the conformal transformation $w=R\ln z$. Combining this
transformation with (\ref{twopoint}) and (\ref{tensortran}), one
obtains the two-point function in the strip geometry:
\begin{equation}
\label{striptwo}
\langle \phi(w_1,{\bar w}_1)\phi(w_2,{\bar w}_2)\rangle=\left[2R\sinh\left(w_{12}/2R\right)\right]^{-2h}\left[2R\sinh\left({\bar
w}_{12}/2R\right)\right]^{-2{\bar h}}.
\end{equation}
Introducing coordinates ($\tau$,$\sigma$) along and across the strip
respectively ($w=\tau+i\sigma$, $\bar w=\tau-i\sigma$;
$-\infty<\tau<\infty$, $0<\sigma<2\pi R$ ), one may expand
(\ref{striptwo}) in the following manner,
\begin{equation}
\label{tausigexp}
\langle\phi(\tau_1,\sigma_1)\phi(\tau_2,\sigma_2)
\rangle=\sum_{n=0}^\infty\sum_{m=-\infty}^{\infty}C_{nm}\,e^{-(h+\bar h
+n)|\tau_{12}|/R}\,e^{-i(h-\bar h + m)|\sigma_{12}|/R}
\end{equation} 
One may also obtain the two-point function in the operator formalism,
leading to the Lehmann expansion:
\begin{equation}
\label{lehmann}
\langle\phi(\tau_1,\sigma_1)\phi(\tau_2,\sigma_2)
\rangle=\sum_{\alpha}|\langle 0|\hat\phi|\alpha\rangle|^2\,e^{-E_\alpha|\tau_{12}|-iP_\alpha|\sigma_{12}|}
\end{equation}
where $E_\alpha$ and $P_\alpha$ are the eigenvalues of the
Hamiltonian and the momentum operator respectively, in the state
$|\alpha\rangle$. Comparing (\ref{tausigexp}) and (\ref{lehmann}) one obtains a relationship
between the eigenvalues of the Hamiltonian and the momentum operator, and
the scaling dimensions in the corresponding CFT:
\begin{equation}
\label{emdim}
E_{\alpha}=\frac{h+\bar h+n}{R},\quad P_{\alpha}=\frac{h-\bar h+m}{R}.
\end{equation}
Restricting our attention
to fields with $h=\bar h=d/2$, one may rewrite
(\ref{striptwo}) in the form
\begin{equation}
\langle\phi(\tau_1,\sigma_1)\phi(\tau_2,\sigma_2)\rangle=R^{-2d}\left[2\cosh\left(\tau_{12}/R\right)-2\cos
\left(\sigma_{12}/R\right)\right]^{-d}.
\end{equation}
One observes that for $\tau_{12}\gg R$ the asymptotic form of the
correlation function is {\em independent} of $\sigma$,
\begin{equation}
\label{decay}
\langle\phi(\tau_1,\sigma_1)\phi(\tau_2,\sigma_2)\rangle\sim
R^{-2d}\exp\left(-d\tau_{12}/R\right)\quad (\tau_{12}\gg R)
\end{equation}
and in this limit one should set $n=m=0$ in equations
(\ref{tausigexp}) and (\ref{emdim}). 

Let us now place the model (\ref{wznw0}) on a thin strip of width $2\pi R$, and
neglect any $\sigma$-dependence of the fields $X_a$. From our considerations in
the previous paragraph, such a procedure preserves the (large $\tau$) asymptotics of the correlation functions.  The action (\ref{wznw0}) becomes
\begin{equation}
S=2\pi R \int d\tau\, G_{ab}[X]\partial_\tau X^a\partial_\tau X^b,
\end{equation}
which may be recognised as the action for a free, non-relativistic
particle, of mass $m=4\pi R$. The corresponding Hamiltonian is the
Laplace--Beltrami operator (multiplied by $-1/2m$),
\begin{equation}
\label{laplace}
\hat H=\frac{-1}{8\pi R\sqrt{G}}\frac{\partial}{\partial
X^a}\left(\sqrt{G} G^{ab}\frac{\partial}{\partial
X^b}\right)
\end{equation}
As we have already established, the eigenvalues of this Hamiltonian
are related to the spectrum of scaling dimensions in our CFT by
equation (\ref{emdim}). Moreover, solution of this Schr{\"o}dinger
equation allows one to obtain explicit expressions for the
eigenstates, $|\alpha\rangle$, appearing in (\ref{lehmann}). 

\subsection{Emergence of the  SL($2,{\Bbb C}$)/SU($2$) symmetry}

 We now wish to establish contact with the one-dimensional
calculations based on the original model (\ref{pruisken}). As is well
established, the scaling trajectories for different values of
$\sigma_{xy}^0$ only start to deviate at the scale $\xi \gg l$ (the
mean free path). For  the sigma model this scale is in the 
deep  infrared, whereas  for the (unknown) critical theory 
 this scale serves as an ultraviolet cut-off. The critical theory is
conformally invariant, and one may map it to the strip in the manner
described above; we assume that the general form of such  expressions holds
 for strips as narrow  as $R
\sim \xi$. Similarly, we assume that the results \cite{Mirlin:onedim}
hold for strips as wide as $\xi$. 
  
 On the one hand, the functions (\ref{bessel}) entering into the
matrix elements (\ref{insight}) should be eigenfunctions of the
Laplace--Beltrami operator for the critical model we seek. On the
other hand, as we are going to show, these functions are solutions of  
  the eigenvalue problem for the Laplace--Beltrami
operator on the manifold SL($2,{\Bbb C}$)/SU($2$). This is striking
because the SL($2,{\Bbb C}$)  symmetry was not the classical
 symmetry  of the
original sigma model and  in fact appears due to the elaborate limit $\eta
\rightarrow 0$ described in the previous sections. 

 Let us describe the details. An arbitrary element $h\in$ SL($2,{\Bbb C}$)/SU($2$) admits the
following decomposition (see equation (18) of \cite{Caux:disferm} and make the
trivial replacements
$(\phi,\mu_+,\mu_-,)\rightarrow(\theta,\mu,\mu^{\ast})$),
\begin{equation}
\label{sl2su2param}
h=\begin{pmatrix}1 & 0 \\ \mu^\ast & 1\end{pmatrix}\begin{pmatrix}e^{\theta} & 0 \\ 0 &
e^{-\theta}\end{pmatrix}\begin{pmatrix}1 & \mu \\ 0 & 1\end{pmatrix}
\end{equation}
where $\theta\in {\Bbb R}$, $\mu\in {\Bbb C}$, and $\mu^\ast$ is the
complex conjugate of $\mu$. Adopting this parameterization, one may write the SL($2,{\Bbb C}$)/SU($2$) WZNW action in the
following form (see equation (19) of \cite{Caux:disferm}),
\begin{equation}
\label{slsuaction}
S=\frac{k}{4\pi}\int d^2x \left[4\partial \theta{\bar\partial}\theta+e^{2\theta}\partial\mu{\bar\partial}\mu^{\ast}\right].
\end{equation}
One may now read off the corresponding metric $G_{ab}$
(c.f. (\ref{wznw0})) and deduce the form of the Laplace--Beltrami
operator (\ref{laplace}) on the SL($2,{\Bbb
C}$)/SU($2$) symmetric space. The resulting eigenvalue equation reads
($X = \exp(- \theta)$):
\begin{equation}
\label{sleigen}
-\left(
X^2\frac{\partial^2}{\partial X^2}+
X\frac{\partial^2}{\partial\mu^\ast\partial\mu}\right)F_{\lambda}(\mu,
\mu^\ast, X)=\lambda F_{\lambda}(\mu, \mu^\ast, X). 
\end{equation}
 The functions $W_j$  satisfying 
equation (\ref{W213})  are related to the following eigenfunctions of
(\ref{sleigen}):
\begin{equation}
F_j = \exp[ik\mu^* + ik^*\mu] W_j(X/|k|) \label{repss}
\end{equation}
 with eigenvalues $\lambda=-j(j+1)$. At the same time the
$\mu$-independent solutions of Eq.(\ref{sleigen}) represented by 
the functions $X^{-j}$  have  the same eigenvalues. We
suggest that these functions with $j = -q$ represent $[\rho]^q$
operators. The  matrix element (\ref{insight})  is then just a
Clebsh-Gordan coefficient.

These  facts
establish a relationship between the SL($2,{\Bbb
C}$)/SU($2$) WZNW theory and the results of the one-dimensional calculations
described in section II. The functions $W_j$  appearing in the 
matrix elements (\ref{insight}) are related to the eigenfunctions
$F_j$ of the
Hamiltonian (\ref{sleigen}). Both the vacuum state $j = 0$ 
and the excited states (the states with $j =
-1/2 + ip/2$) belong to the coherent state representations  of the SL($2,{\Bbb
R}$) group (see Appendix \ref{SLrep} and  \cite{Carmeli} for details). 
The representations of the 
excited states are distinct in the respect that their angular momentum
is a complex number with continuously varying imaginary part. The
scaling dimension of the vacuum state is zero whereas those of the excited states are 
proportional to $\lambda = (1+p^2)/4$. They coincide with the exponents  in equation
(\ref{contin}). The constant wave function, though being a solution of
equation (\ref{sleigen}), is not orthogonal to the eigenfunctions and
therefore does not belong to the basis of eigenstates. In the subsequent sections we shall study the SL($2,{\Bbb C}$)/SU(2)
WZNW model in more detail, and show how it may emerge as an
independent subsector of some `grand' (supergroup manifold) theory.

 The appearance of the SL($2,{\Bbb C}$)/SU(2)
 symmetry  absent in the original sigma model where it is explicitly 
broken at
 finite frequencies 
  is a highly non-trivial fact and deserves comment. The calculations
 of the local DOS performed in \cite{Mirlin:onedim,MF} employ  the
 general regularization procedure described in Section II. Taking the
 limit $\eta \rightarrow 0$ and keeping $(\eta Q)$ constant involves
 working in the limit of very large values of fields. In this limit
 the $\eta$-term (\ref{eta})  in the action becomes of order of unity 
 (such that $\eta$-dependence disappears) 
and contributes the term linear in $X$ to the Schr\"odinger equation (\ref{W213}).   One can say that this
 term is some sort of ``quantum anomaly''.

 Thus, in the original sigma model 
the term linear in $X$ is generated as a
potential contribution. 
On the other hand, as we have seen from equations 
(\ref{sleigen}) and (\ref{repss}),
  equation (\ref{W213}) can be interpreted  as a subsector of 
 the pure Laplace-Beltrami-type  equation (\ref{sleigen}), which
contains  no potential terms. This subsector specifies 
 representations of the particular type (\ref{repss}). 
It is important that the representations of $[\rho]^q$ (that is
 the discrete series) are solutions of the same equation
 (\ref{sleigen}), but in the $\mu$-independent subsector.  This fact
allows one to consider (\ref{repss}) and $[\rho]^q$ as 
 representations of the same group. Thus the 
 one-dimensional limit has a {\it
hidden}  dynamical symmetry and this fact gives another argument in
favour of our  conjecture that the critical point possesses  the 
SL($2,{\Bbb C}$)/SU(2) symmetry.

\section{Gauged WZNW models} 
\label{sect:gauged}
In this section we consider the WZNW model on the supergroup manifold
PSL($2|2$)=SL($2|2$)/GL($1$) (see Appendix \ref{sect:super} for more
details on supergroups). These
two-parameter models are quite remarkable in
that they are believed to be conformal along (marginal) lines of parameter
space, rather than isolated points
\cite{Berkovits:ramramflux,Bershadsky:psl}. Zirnbauer has proposed a theory
describing the plateau transitions in a region of this parameter space
not endowed with the Ka{\v c}--Moody symmetry
\cite{Zirnbauer:integer}. For the purposes of this paper we consider the model where
the Ka{\v c}--Moody symmetry is present (the `Ka{\v c}--Moody'
point). In particular, in section \ref{sect:decoup}, we
demonstrate that the bosonic and fermionic sectors {\em decouple}; the
non-compact bosonic sector is described by the SL($2,{\Bbb C}$)/SU(2)
WZNW model. 

 The standard Goddard--Kent--Olive (GKO) procedure \cite{Goddard:coset} for
dealing with such coset spaces may be formulated as a gauged 
WZNW model (for more details see \cite{Fransesco:cft, Ketov:cft,
Tsvelik:boson}). The equations for correlation functions can be
obtained by the gauge dressing of the conventional Knizhnik--Zamolodchikov
equations 
 \cite{Kogan:gauged}.  We consider the WZNW model on the SL($2|2$)
group first and then gauge away the GL($1$) subsector. In order to make our arguments transparent, we discuss some
general aspects of the theory of WZNW models. 

A convenient starting point for discussing the coset $G/H$, is the
(Euclidean) WZNW action on the supergroup manifold $G$,
\begin{equation}
\label{WZNW}
{\hat S}[g]={\hat S}_0[g]+k{\hat \Gamma}[g],
\end{equation}
where
\begin{eqnarray}
{\hat S}_0[g] & = & \frac{1}{4\lambda^2}\int d^2 \xi\, \str\left(\partial^\mu g^{-1}\,\partial_\mu g\right)\\
{\hat \Gamma}[g] & = & \frac{-i}{24\pi}\int d^3 x \,\epsilon^{\mu\nu\rho}\,\str\left(\ginv\parmu g\,\ginv\parnu g\,\ginv\parrho g\right).
\end{eqnarray}
The matrix field $g(\xi)$ is taken to be an element of some
supergroup $G$, the hats indicate the presence of the
supertrace\footnote{In the subsequent analysis removal of the hats in
any term corresponds to the replacement of supertrace by ordinary
trace. Our conventions for supertrace (and superdeterminant) are
defined in Appendix \ref{sect:super}.}, $\xi^\mu=(\xi^1,\xi^2)$ are the coordinates of our
two-dimensional (Euclidean) space, $\lambda^2$ and $k$ are
dimensionless parameters ($\hbar=1$). The Wess--Zumino term,
${\hat \Gamma}[g]$, is defined by the integral over the
three-dimensional ball with coordinates $x^\alpha$; the boundary being
identified with our two-dimensional space. Since the Wess--Zumino term
contributes to the equations of motion (see below), and hence to the
perturbative beta function, it is {\em not} to be confused with the
topological term.  

The action (\ref{WZNW}) satisfies the Polyakov--Wiegmann 
identity\footnote{In the proof of the Polyakov--Wiegmann identity we
exploit the cyclic property of the trace. For supermatrices, it is the
supertrace which has this property (see Appendix \ref{sect:super}). The form of the Polyakov--Wiegmann
identity for supermatrices is the same as that for ordinary matrices,
providing we replace trace by supertrace. }\cite{Poly:multi},
\begin{equation}
\label{PW}
{\hat S}[ab]={\hat S}[a]+{\hat S}[b]+\int d^2 \xi\,
\omega^{\mu\nu}\,\str(a^{-1}\parmu a \,b \parnu b^{-1});\quad\omega^{\mu\nu}=\frac{\delta^{\mu\nu}}{2\lambda^2}-\frac{ik\epsilon^{\mu\nu}}{8\pi}.
\end{equation}
The classical equation of motion follows from the requirement that the
action be stationary ($0=\delta S\equiv S[g+\delta g]-S[g]$) under the
replacement $g\rightarrow g+\delta g\equiv g(1+g^{-1}\delta g)$. Substituting the latter form of the variation into (\ref{PW}), and keeping terms of order $\delta g$, 
\begin{equation}
0=\delta S=-\int d^2\xi\,\omega^{\mu\nu}\,\str(g^{-1}\partial_\mu g\,\partial_\nu(g^{-1}\delta g))\Rightarrow\partial_\nu(\omega^{\mu\nu}g^{-1}\partial_\mu g)=0.
\end{equation}
Thus for $\lambda^2=4\pi/k$ the field equation becomes
\begin{equation}
\label{currents}
\partial(g^{-1}\bar\partial g)=0\Longleftrightarrow \bar\partial(\partial
g g^{-1})=0.
\end{equation}
We shall denote the action $\hat S[g]$ at the Ka{\v c}--Moody point
($\lambda^2=4\pi/k$) by the symbol $\hat W[g]$:
\begin{equation}
\label{KMWNZM}
{\hat W}[g]\equiv k\hat I[g]=\frac{k}{16\pi}\int d^2\xi\,\str(\partial^\mu g^{-1}\partial_\mu g)+k{\hat\Gamma}[g]. 
\end{equation}
The action (\ref{KMWNZM}) is invariant under the semi-local transformation
\begin{equation}
g(z,\bar z)\rightarrow \Omega(z)g(z,\bar z){\bar \Omega}^{-1}(\bar z)
\end{equation}
where $\Omega(z)$ and $\bar \Omega(\bar z)$ are arbitrary elements of
$G$. This invariance is made manifest by the Polyakov--Wiegmann
identity (\ref{PW}) with $\lambda^2=4\pi/k$,
\begin{equation}
\label{PW1}
{\hat W}[ab]={\hat W}[a]+{\hat W}[b]+\frac{k}{2\pi}\int d^2\xi\,\str(a^{-1}\bar\partial a\,b\partial b^{-1}).
\end{equation} 
One may consider promoting
this semi-local symmetry to a true local symmetry, with the introduction of
auxiliary {\em gauge fields}. In particular, the action
\begin{equation}
\label{gauged}
{\hat W}[g,h,{\tilde h}]={\hat W}[h^{-1}g {\tilde h}]-{\hat W}[h^{-1}{\tilde h}]
\end{equation}
is clearly invariant under the combined (gauge) transformations
\begin{equation}
\label{gautrans}
g \rightarrow \lambda(z,{\bar z}) g\lambda^{-1}(z,{\bar z}),\quad
h\rightarrow \lambda(z,{\bar z}) h, \quad
\tilde h\rightarrow  \lambda(z,{\bar z}) \tilde h.
\end{equation}
Applying the Polyakov--Wiegmann identity (\ref{PW1}) and defining the following {\em gauge fields}
\begin{equation}
\label{gaufields}
A=\tilde h\partial {\tilde h}^{-1}, \quad \bar A= h\bar\partial h^{-1}
\end{equation}
one may rewrite the action (\ref{gauged}) in the following form,
\begin{eqnarray}
\label{gauarray}
{\hat W}[g,A,\bar A] & = & {\hat W}[g]+\frac{k}{2\pi}\int d^2\xi\,\str\left(Ag^{-1}\bar\partial
g-\bar A\partial gg^{-1}+Ag^{-1}\bar A g-A\bar A\right)\\
& = &  {\hat W}[g]+\frac{k}{2\pi}\int d^2\xi\,\str\left(A{\bar
J}_A+{\bar A}J_A\right)
\end{eqnarray}
where we have introduced the gauge invariant generalizations of the
conserved currents (\ref{currents}):
\begin{equation}
{\bar J}_A = g^{-1}(\bar\partial +{\bar A})g, \quad J_A=-(\partial + A) gg^{-1}.
\end{equation}
The gauge fields $A$ and ${\bar A}$ are non-propagating, and play the
r{\^ o}le of Lagrange multipliers forcing certain currents to vanish. Choosing the gauge fields so that the currents associated with
the group $H$ are set to zero, we may describe the
coset space $G/H$. The action (\ref{gauarray})
will be our staring point in the next section. 
\section{Decoupling at the Ka{\v c}--Moody point}
\label{sect:decoup}
In order to see how the SL($2,{\Bbb C}$)/SU(2) WZNW may emerge as an
independent subsector of the Zirnbauer model (at the Ka{\v c}--Moody
point) or some other `grand' supermanifold theory, we parametrize the
supergroup manifold using the Gauss decomposition. The use of such decompositions is well established
in the the free-field approach to WZNW models (see, for example 
\cite{Geras:free} and references therein). In
particular, such an approach has been followed in the study of the gl($n|n$) current
algebras and associated topological field theories
\cite{Isidro:gltop}.
We consider the gauged WZNW model\footnote{Throughout this
section $\int\equiv \int\, d^2\xi$.},
\begin{equation}
\label{gaugy}
 {\hat W}[g,A,\bar A] = {\hat W}[g]+\frac{k}{2\pi}\int\,\str\left(Ag^{-1}\bar\partial
g-\bar A\partial gg^{-1}+Ag^{-1}\bar A g-A\bar A\right)
\end{equation}
in which the supergroup element, $g\in$ SL($2|2$),
admits the following Gauss decomposition (see equation \ref{slgauss}) 
\begin{equation}
g=e^\Phi\gamma;\quad \gamma=\begin{pmatrix}{\openone} & 0 \\ \lambda & {\openone}\end{pmatrix}\begin{pmatrix}a &0 \\ 0 & b\end{pmatrix}\begin{pmatrix}{\openone} & \chi \\ 0 & {\openone}\end{pmatrix}
\end{equation}
where $\Phi\in{\Bbb C}$, $\lambda$ and $\chi$ are arbitrary $2\times
2$ Grassmann-odd matrices, and $a$ and $b$ are arbitrary $2\times 2$ {\em unimodular}
matrices (${\rm det}(a)={\rm det}(b)=1$) with Grassmann-even
entries. One may recover the Zirnbauer's `base manifold' by taking $a\in\,$
SL($2,{\Bbb C}$)/SU(2) and $b\in\,$ SU(2). In order to set the gl($1$) currents equal
to zero, and thus describe the PSL($2|2$) coset, we choose (see equation \ref{isolate}) 
\begin{equation}
A=\frac{\mu}{2}\Sigma, \quad {\bar A}=\frac{\bar\mu}{2}\Sigma; \quad
\Sigma=\begin{pmatrix}{\openone} & 0 \\ 0 & {-\openone}\end{pmatrix}
\end{equation}
Applying the Polyakov--Wiegmann identity (\ref{PW1}) to the first term
of (\ref{gaugy}),
\begin{equation}
{\hat W}[g]= {\hat W}[e^{\Phi}]+{\hat
W}[\gamma]+\frac{k}{2\pi}\int\,
{\bar\partial}\Phi\,\str\left(\gamma\partial\gamma^{-1}\right) ={\hat W}[\gamma]
\end{equation}
where we have used the fact that ${\hat W}[e^{\Phi}]=0$ and that
$\str\left(\gamma\partial\gamma^{-1}\right)=0$. Repeated application
of the Polyakov--Wiegmann identity (\ref{PW1}) shows that
\begin{equation}
{\hat W}[\gamma]=W[a]-W[b]+\frac{k}{2\pi}\int\,\tr(b^{-1}{\bar\partial}\lambda a\partial \chi).
\end{equation}
Using the fact that ${\str(\Sigma\cdots)}={\tr(\cdots)}$ for
arbitrary arguments, the second
term in (\ref{gaugy}) may be written
\begin{equation}
\frac{k}{2\pi}\int\,\left[\tr\left(\frac{\mu}{2}\gamma^{-1}\bar\partial
\gamma-\frac{\bar\mu}{2}\partial \gamma\gamma^{-1}+\frac{\mu\bar\mu}{4}\gamma^{-1}\Sigma \gamma\right)+2\mu\bar\partial\Phi-2\bar\mu\partial\Phi\right].
\end{equation}
Straightforward matrix manipulation of $\gamma$, together with the
fact that $\tr(a^{-1}\bar\partial a)=0$, $\tr(\partial a a^{-1})=0$, and
likewise for $b$, yields
\begin{eqnarray}
\tr(\gamma^{-1}\bar\partial\gamma)& = & 2\tr\left(b^{-1}\bar\partial\lambda a\chi\right),\\
\tr(\partial\gamma\gamma^{-1})& = & 2\tr\left(b^{-1}\lambda a\partial\chi\right),\\ 
\tr(\gamma^{-1}\Sigma\gamma)& = &-4\tr\left(b^{-1}\lambda a\chi\right).
\end{eqnarray} 
Combining the above results we may write
\begin{equation}
{\hat W}[g,A,\bar A]=W[a]-W[b]+\frac{k}{2\pi}\int\,\left[
\tr\left(b^{-1}(\bar\partial-\bar\mu)\lambda
a(\partial+\mu)\chi\right)+2\mu\bar\partial\Phi-2\bar\mu\partial\Phi\right].
\end{equation}
It is convenient to introduce new fermionic fields
$\lambda^\prime$ and $\chi^\prime$ via 
\begin{equation}
\lambda=b\lambda^\prime,\quad \chi=a^{-1}\chi^\prime.  
\end{equation}
Since $a$ and $b$ are unimodular matrices, the corresponding Jacobian is unity. The trace term becomes 
\begin{equation}
\tr\left[\left(\bar\partial-\bar\mu+b^{-1}\bar\partial
b\right)\lambda^{\prime}\left(\partial+\mu+a\partial
a^{-1}\right)\chi^\prime\right] =\tr \left[(\bar\partial +\bar
A^b)\lambda^\prime(\partial +A^a)\chi^\prime\right]
\end{equation} 
in which we have reparametrized, $\mu=\partial\alpha$, and
$\bar\mu=\bar\partial\beta$, and introduced the gauge fields
\begin{equation}
\bar A^b=(e^\beta b^{-1})\bar\partial(e^{-\beta}b), \quad
A^a=(e^{-\alpha}a)\partial(e^\alpha a^{-1}).
\end{equation} 
We make a further change of fermionic variables 
\begin{equation}
\lambda^{\prime\prime}=\left(\bar\partial+\bar
A^b\right)\lambda^{\prime},\quad
\chi^{\prime\prime}=\left(\partial+A^a\right)\chi^\prime  
\end{equation}
and note that one may express the corresponding Jacobian in terms of the WZNW action on the matrices $a$ and $b$ (not $b^{-1}$) respectively
 \cite{Poly:nonab,Poly:multi}.  Recalling that the fermionic fields
are 2$\times$2 matrices, this leads to a level shift of 2 in the WZNW
models defined over $a$ and $b$. Such level shifts were also encountered in
\cite{Isidro:gltop,Berkovits:ramramflux}. One obtains the following action for
the PSL$(2|2)$ WZNW model,
\begin{equation}
k{\hat {\rm I}}[{\rm PSL}(2|2)] = (2+k){\rm I}[a] + (2-k){\rm I}[b] +\cdots 
\end{equation}
in which the ellipsis indicates terms {\em independent} of $a$ and $b$;
these include the contributions of the bosonic fields $\alpha$ and $\beta$, the fermions
$\lambda^{\prime\prime}$ and $\chi^{\prime\prime}$ and ghosts.
In the case of the Zirnbauer `base manifold', $a\in$SL($2,{\Bbb C}$)/SU(2)
and $b\in$SU(2). The level
shifts are important here, and the resulting contribution to the central charge
from these (renormalized) WZNW models is level-independent:
\begin{equation}
C = \frac{3(k + 2)}{(k + 2) - 2} + \frac{3(k - 2)}{(k - 2) + 2}=6
\end{equation}
In particular, one observes that the SL($2,{\Bbb C}$)/SU(2) WZNW model
emerges as an {\em independent} subsector of the theory.

\section{Correlation Functions of Local Density of States}
\label{sect:dos}
We have provided evidence to suggest that the correlation functions of
the {\it normalized} local DOS may be obtained from the {\em critical} SL($2,{\Bbb C}$)/SU(2)
WZNW model, in which the level $k$ is fixed by the known scaling 
dimensions (\ref{scalrel}). This model has
been studied before, even in the context of the theory of disorder
\cite{Caux:disferm}. From the preceding sections, we have seen how
this theory may be embedded, for example, into a larger theory such as
the PSL($2|2$) model (at the Ka{\v c}--Moody point). Taking into
account the level shifts induced in this embedding
one may write the
action for the SL($2,{\Bbb C}$)/SU(2) WZNW model (\ref{slsuaction}) in the following form
\begin{equation}
\label{slaction}
S=\frac{(k+2)}{4\pi}\int d^2x \left[4\partial
\theta{\bar\partial}\theta+e^{2\theta}\partial\mu{\bar\partial}\mu^{\ast}\right]
\end{equation}
A comparison between the decomposition of SL($2,{\Bbb C}$)/SU(2)
appearing in (\ref{sl2su2param}) and
the Gauss decomposition of SL($2,{\Bbb R}$) 
\begin{equation}
g=\begin{pmatrix}1 & 0 \\ \gamma & 1\end{pmatrix}\begin{pmatrix}e^{\theta} & 0 \\ 0 &
e^{-\theta}\end{pmatrix}\begin{pmatrix}1 & \psi \\ 0 & 1\end{pmatrix}
\end{equation}
where $\gamma, \theta, \phi\in{\Bbb R}$ is convenient at this
stage. In particular, one may obtain
the parametrization of SL($2,{\Bbb C}$)/SU(2) from that of SL($2,{\Bbb R}$), by first complexifying the real parameters
$\gamma$ and $\psi$, and then making the identification
$\gamma^\ast=\psi$. In this way, one may obtain
the SL($2,{\Bbb C}$)/SU(2) correlation functions from the (much
studied) SL($2,{\Bbb R}$) WZNW model\footnote{Despite the fact that 
the SL($2,{\Bbb R}$)
model cannot be defined as an Euclidean  path integral, it can be
defined algebraically. Following this procedure one can derive the
Knizhnik-
Zamolodchikov equation for the correlation functions. This equation
 is not different from the corresponding equation for the SL($2,{\Bbb
C}$)/SU(2) model.} one treats $\gamma$ and $\psi$
as {\em independent real} fields and performs the appropriate continuation wherever necessary. 

The action (\ref{slaction}), is characterized by a single coupling
constant $k$, which determines the scaling dimensions:
\begin{equation}
\label{conf}
d_{j}=-2j(j+1)/k
\end{equation}
There are no restrictions on the value of $k$, and the values of $j$
are related to representations of the SL(2,${\Bbb R})$ group (see
Appendix \ref{SLrep}).

As we have discussed in Section \ref{dicorep}, the available information
about the local DOS appears to be self-contradictory. On the one hand, the fusion rules for $[\rho({\bf r})]^q$ suggest that these 
are operators belonging to the sl(2,${\Bbb R}$) representation with $j = -
q$ and with scaling dimensions given by (\ref{conf}). On the other hand, the
two-point correlation function in the strip geometry is given by
an integral over {\em continuous} representations of sl(2,${\Bbb R}$),
with $j = -1/2 + ip/2$. As we have demonstrated in Section \ref{dicorep},
this paradox is resolved if, in fact, the two-point function of the local DOS is given as the limit of a
{\em four-point} function in our field theory:
\begin{equation}
G_q(1,2) = \lim_{|r_3| = 0, |r_4| \rightarrow \infty}\langle[\rho^q](1)[\rho^q](2)\Psi_0(3)\Psi_0(4)\rangle \label{cor} 
\end{equation}

The four-point correlation function appearing in this relation (which
we shall subsequently denote by $G_q(1,2,3,4)$) is determined by the Knizhnik--Zamolodchikov
equations for the SL(2,${\Bbb R}$) WZNW model; these equations have
been derived in \cite{FZ, Kogan:gauged}. The operators in (\ref{cor})
are matrix elements of infinite-dimensional matrices realizing
particular representations of the group (information about
representation theory of sl(2,${\Bbb R}$) is given in Appendix
4). Instead of working with these particular matrix elements, it is
more convenient to calculate the correlation function between {\it all
possible} tensors belonging to representations with $j = 0, j =
-q$. We emphasise that one does not need to worry about exact
definition of $\Psi_0$ because the Knizhnik--Zamolodchikov
equations together with the asymptotics of the solution 
{\it automatically select} the necessary primary fields.  By that token
$\Psi_0$ standing in the correlation function is {\it the } very 
primary field with zero scaling dimension we need. 
 It is necessary to mention here that there may be several 
fields with zero dimensions  with different symmetry properties 
 and unusual OPE \cite{KT} 
 leading to other solutions of  Knizhnik--Zamolodchikov equations 
 for  correlation function (\ref{cor}). However these solutions are
 excluded since they do not 
lead to the desirable quasi-one-dimensional asymptotics.

 We introduce the fields $\rho^q(z,y),\Psi_0(z,y)$ as
described in Appendix 4. Now each point in (\ref{cor}) is characterized
by two complex coordinates $z$ and $y$. Invariance under the projective
transformations of the plane\footnote{Namely transformations of the
form $w = (az + b)/(cz + d)$, where $ad - bc = 1$.}, together with the sl(2,${\Bbb
R}$) Ward identities, restricts the four-point function to have the form,  
\begin{equation}
\label{fourish}
G_q(1,2,3,4) = \frac{1}{|z_{12}|^{4h_q}|y_{12}|^{4q}}{\cal F}_q(z,\bar z; t,
  \bar t); \quad \quad z = \frac{z_{32}z_{41}}{z_{31}z_{42}}, \quad t = \frac{y_{32}y_{41}}{y_{31}y_{42}},
\end{equation}
where we have used the fact that $[\rho^q]$ has conformal
dimension $h_q (= d_q/2)$, and sl(2,${\Bbb R})$ spin $j=-q$, whereas $\Psi_0$ has zero conformal dimension and $j=0$. 
By definition, the function (\ref{fourish}) is invariant under the
permutation of $z_1$ and $z_2$ which implies
\begin{equation}
{\cal F}_q(z, \bar z; t, \bar t) = {\cal F}_q(1/z,1/\bar z,1/t,1/\bar t). \label{crossish}
\end{equation}
In the limit, $z_{12} \rightarrow 0$, corresponding to the fusion of the
local DOS, one obtains
\begin{eqnarray}
\langle[\rho^q](1)[\rho^q](2)\Psi_0(3)\Psi_0(4)\rangle & \rightarrow & |z_{12}|^{-4h_q +
2h_{2q}}\,\langle[\rho^{2q}](2)\Psi_0(3)\Psi_0(4)\rangle \nonumber\\
& = &  |z_{12}|^{-4h_q +2h_{2q}} \,C_{2q}^{00}\,|z_3 - z_2|^{- 2h_{2q}},
\end{eqnarray}
where $C_{2q}^{00}$ is a structure constant of the conformal field
theory. 
This 
 fixes the asymptotics of ${\cal F}_q(z)$ at $1-z,1- y \rightarrow 0$:
\begin{equation}
{\cal F}_q(z,y) \sim (1 - z)^{h_{2q}}y^{2q} \label{asympt2}
\end{equation}
We may explore different asymptotics of ${\cal F}_q(z)$ in the strip geometry. Under the transformation
$z = \exp(w/R)$ the point 3 goes to $-\infty$ and the point 4 goes to
$+\infty$. Considering only the holomorphic part and omitting the
$y$-dependence, one obtains
\begin{equation}
G_q(1,2,3,4) = [2R\sinh(w_{12}/2R)]^{-2h_q}{\cal
 F}_q\left[\exp(w_{12}/R)\right] \label{strip}
\end{equation}
Thus in the strip geometry the correlation function is translationally
 invariant. The limit ${\rm Re}\, w_{12} \gg R$  is governed by the
 asymptotics  
of ${\cal F}_q(z)$ as $z \rightarrow \infty$ and the limit ${\rm Re}\,
 w_{21} \gg R$ is governed by the asymptotics as  $z \rightarrow 0$. 
Both limits are related by the crossing invariance
condition (\ref {crossish}). 

 Thus, the fusion rule (\ref{discfusion}) and the expansion in
continuous representations (\ref{corrfun}) appear in {\em different
channels} of the four-point function, thus resolving the apparent paradox. 

According to \cite{Kogan:gauged, FZ} the Knizhnik--Zamolodchikov
equation for model (\ref{slaction}) is given by
\begin{equation}
\left\{ k\frac{\p}{\p z_i} - \sum_{j \neq i}\frac{1}{z_{ij}}\left[(y_{ij})^2\frac{\p^2}{\p y_i\p y_j} + 2(y_{ij})\left(j_j\frac{\p}{\p y_i} - j_i\frac{\p}{\p y_j}\right) -
 2j_ij_j\right]\right\}G(1\cdots 4) = 0
\end{equation}
as may be seen by combining equations (\ref{sl2rgen}) and (\ref{kz})
together with a suitable redefinition of $k$. Substituting (\ref{fourish}) into this equation, and using the
global conformal invariance to map
$(z_1,z_2,z_3,z_4)\rightarrow(1,z,0,\infty)$ (and similarly for $y$ using the
sl(2,${\Bbb R}$) Ward identities)
one arrives at the following partial differential equation for  ${\cal F}(z,t)$


\begin{equation}
\p^2_{t}{\cal F}\frac{(t - z)t(t - 1)}{z(z - 1)} + \p_t{\cal
F}\left[\frac{(1 - t)^2}{z - 1} + \frac{t(2q - t)}{z}\right] +
 k\p_z{\cal F} = 0 \label{Eq}
\end{equation}
>From equation (\ref{asympt2}) we conclude that in the `discrete' channel $|z - 1| \ll 1$, ${\cal F} 
=  (z - 1)^{h_{2q}}f(t)$. Equation (\ref{Eq}) may now be written as
\begin{equation}
(1 - t)^2\p_t[t\p_t f] + kh_{2q}f = 0.\label{eq}
\end{equation}
The solution of this equation takes the form
\begin{eqnarray}
f[t = 1/(1 - x)]& =& F(2q, 1-2q, 1;x),\nonumber \\ 
                & =& [\p_x]^{2q - 1}\left[x^{4q
- 1}(1 - x)^{-1}\right],\nonumber\\
&=& (2q - 1)!(t^{-1} - 1)^{2q} + \sum_{k = 0}^{2q - 1}(1 -
t^{-1})^{-k}k!C^{k + 2q}_{4q - k}.
\end{eqnarray}
Since equation (\ref{eq}) is invariant under the transformation  $t \rightarrow
1/t$ the second linearly independent solution is 
\begin{equation}
\tilde f(t) = f(1/t)
\end{equation}
and the entire crossing invariant solution at $|z - 1| <<1$ is given by 
\begin{equation}
{\cal F}(z,\bar z;t, \bar t) = |1 -z|^{2h_{2q}}\{A[f(t)f(\bar t) +
f(1/t)f(1/\bar t)] + B[f(t)f(1/\bar t) + f(\bar t)f(1/t)]\} \label{zet1}
\end{equation}

Following equation (\ref{asympt1}) at  $|z| \ll  1$ (the `continuous' channel)
 we look for the solution in the form of a linear combination 
of power law solutions:
\begin{equation}
{\cal F}  = \int d\mu(p)|z|^{2D(p)/k}F_p(t,\bar t) \label{integral}
\end{equation}
where
\begin{equation}
D(p) = (1 + p^2)/4 - kh_q = (q - 1/2)^2 + p^2/4.
\end{equation}
The function $F_p$ obeys the equation:
\begin{equation}
t^2(1-t)\p^2_tF + t(2q - t)\p_tF +  D(p)F = 0.
\end{equation}
After the change of variables $t = 1/\tau$ one obtains the hypergeometric equation:
\begin{equation}
\tau(1-\tau)\p^2_{\tau}F + [1 - (2 - 2q)\tau]\p_tF -  D(p)F = 0
\end{equation}
There are two independent solutions of this equation:  
\begin{eqnarray}
F^{(1)}_p(t) & = & F(j, j^*, 1; 1/t), \nonumber\\
F^{(2)}_p(t) & = & \ln t\, F(j, j^*, 1; 1/t) + R(1/t), 
\end{eqnarray}
where $j = 1/2 - q + ip/2$ and $R(x)$ is regular as $ x \rightarrow 0$. Thus, the general solution (\ref{integral}) at small $|z|$ is 
\begin{equation}
{\cal F}  = \int d\mu(p)|z|^{2D(p)/k}C_{ab}(p)F^{(a)}_p(t)F^{(b)}_p(\bar t) \label{integral1}
\end{equation}
where the matrix $C_{ab}$ ($a,b = 1,2$) is determined by the crossing
 symmetry (\ref{crossing}).  At the moment we do not have a complete solution of equation (\ref{Eq})
 which prevents us from determining  all the constants in our equations
 and OPEs. It appears, however, that the solution is possible. Here
 we shall just outline the method leaving the detailed analysis for our next publication.

The general solution contains the parameters $y_i$ which replace
 matrix indices for operators belonging to infinite dimensional
 representations. Presumably  the local
 DOS corresponds to certain matrix elements. In order to identify 
 $\rho^q$ it is instructive to take a closer look at the solution on
 the strip. Combining equation (\ref{strip}) with equation (\ref{integral1}) we
 obtain the following expression for the asymptotics of the 
two-point correlation function
 of operators $\rho^q(w,y)$:
\begin{eqnarray}
\overline{\rho^q(w_1,y_1)\rho^q(w_2,y_2)} &=& 
|y_{12}|^{-4q}\int d\mu(p)\exp\left[- (1 + p^2)
|\tau_{12}|/2kR\right]C_{ab}(p)F_a(t)F_b(\bar t)  \label{answer}
\end{eqnarray}
where $w = \tau + i\sigma$ ($0 < \sigma < 2\pi R$). In the limit
 $|\tau_{12}| \gg R$ the prefactor of the integral 
becomes unity. One recovers  the correlation function (\ref{insight}) 
leaving in the expansion of (\ref{answer}) in $t, \bar t$ only 
 the term $t^{- 2q}{\bar t}^{1-2q}$.

 We use this fact to determine which matrix element represents
 $\rho^q$. Therefore let us assume that this is a  matrix element
 of $\rho^q(y)$ containing $y^n$. In order to extract it  from
 the $y$-dependent correlation function one has to integrate it  with the weight $y^{-1 -
 n}$ over a
 contour in the $y$-plane surrounding zero. Then  we have 
\begin{multline}
\int \prod_{i=1}^4dy_i \, y_1^{-1 - n}y_2^{-1 -n}y_{12}^{-2q}{\cal
 F}(t) = \nonumber\\
\int d\xi_1 d\xi_2 (\xi_1\xi_2)^{-(1 + n)}\xi_{12}^{2(1 - q)}\int dt
 \,t^{-2n -2q + 2}{\cal F}(t)\int dx \frac{x}{(1 - x)^2(x - t)^2} \sim 
 \nonumber\\
\int d\tau F(j.j^*,1;\tau)\frac{1 + \tau}{(1 - \tau)^3}\tau^{2(q + n - 1)}
\end{multline}
where we have performed  the transformation 
\[
x = \frac{y_{32}}{y_{31}},\quad \quad  y_4 = \frac{y_1y_{32} - ty_2y_{31}}{y_{32} -
ty_{31}}.
\]
Notice that we integrate over $y_3, y_4$ with unit weight. 
In order to recover the term $\tau^{2q - 1}$ in the expansion of the
hypergeometric function corresponding to the 
necessary matrix element, one has to put  $n = - 2q$. Thus we identify
$\rho^q$ with the {\it highest weight} of the $j = -2q$
representation.

The Knizhnik-Zamolodchikov equation (\ref{Eq}) can be solved by the
method suggested in \cite{FZ}. Let us make the change of variables:
\begin{equation}
{\cal F}(z,t) =t^a(t - 1)^b(t - z)^cz^d(z - 1)^eY(t,z)
\end{equation}
where
\begin{gather}
a = - (1 + k)/2, \quad b = q - (k +1)/2, \quad c = 1/2 - q,\nonumber \\
d = -(2q - 1)/2k,\quad e = (2q - 1)^2/2k. 
\end{gather}
 In terms of $Y$, equation (\ref{Eq}) becomes 
\begin{eqnarray}
-k^{-1}\p^2_t Y 
&=& \frac{z(z -1)}{t(t - 1)(x - t)}\p_z Y + \frac{1 -
2t}{t(t - 1)}\p_t Y + \nonumber\\ &+& [h_1(t - z)^{-2} + h_2t^{-2} +
h_3(t - 1)^{-2} - \kappa[t(t - 1)]^{-1}]Y 
\end{eqnarray}
where 
\begin{equation}
\kappa = h_1 + h_2 +  h_3 - h_4 - (3k + 2)/4
\end{equation}
This equation coincides with the equation 
for the 5-point function of vertex operators 
in the Liouville theory
\begin{equation}
S_L = \int d^2x\left[\frac{k}{4\pi}(\p_{\mu}\phi)^2 + 2(k + 1){\cal
R}\phi + \eta\exp(-2\phi)\right]
\end{equation}
(${\cal R}$ is the Riemann curvature of the world sheet) with the  central charge equal to 
\begin{equation}
\hat C = 1 + 6(\sqrt k + 1/\sqrt k)^2
\end{equation}
The operators of the WZNW model are identified as 
vertex operators of the Liouville theory: 
\begin{equation}
V_{n} = \exp(n\phi)
\end{equation}
with certain $n$s. 

 This establishes a link between the Liouville theory and the theory
 of the local DOS at the plateau transitions in the IQHE. 
\section{Conclusions}
We summarize here the results of this paper:
\begin{enumerate}
\item We have described  a general procedure for calculating
correlation functions of local DOS at the plateau transitions in the
IQHE. This procedure requires  insertions of  additional vacuum operators 
into any correlation function, so as to modify the
ground state.

\item We have provided arguments, that the correlation functions of
the local DOS are appropriately described in terms of the SL($2,{\Bbb
C}$)/SU(2) WZNW model at the usual Ka{\v c}--Moody point and with a suitably
chosen level $k$. The comparison with the numerics gives $6 < k < 8$. 
In this model we have identified the operators
corresponding to the properly normalized local DOS, and calculated their
correlation functions. 

\item We have demonstrated that the SL($2,{\Bbb C}$)/SU(2) WZNW model
may emerge as an independent subsector of some supergroup theory, 
and may indeed describe a disordered system.  
\end{enumerate}

There are many unresolved problems which require further study. 
A number of these are outlined below. 

 The model contains a free parameter $k$  which we fix by comparison 
with the known scaling dimensions. It is not clear at the moment what
mechanism is responsible for
pinning $k$ to this particular value. The example given in Appendix
\ref{BRST} demonstrates the kind of surprises one can expect in dealing
with theories on non-compact manifolds.  

 We have studied in some detail only the correlation functions and scaling
dimensions of the local DOS. We are not in a position to discuss
 the correlation
function exponent $\nu$ which is, presumably, 
 determined by an operator which does not
belong to the non-compact bosonic sector of the theory.  


 Even in the non-compact bosonic sector of the theory  there are important 
unsolved problems.  For example, we
have not yet solved the Knizhnik--Zamolodchikov equation (\ref{Eq})
which determines the two-point correlation function of the local
DOS. In subsequent publications we shall  describe a procedure for calculation of the multi-point correlation functions. 
This will allow us to explain the termination of the multifractal
spectrum. 
 As we have mentioned in Section \ref{subscaldim} the multifractal spectrum of the local
 DOS terminates at $q \sim 2.5$. If  the SL(2,${\Bbb C}$)/SU(2) WZNW model is the
 correct model for the local DOS, it must possess some  termination
 mechanism. This may be a  mechanism similar to the one described in
\cite{Caux:disferm} in the context of  the theory of Dirac
 fermions with gauge potential disorder. The latter mechanism was 
  related to the 
logarithmic nature of the theory.
         
\section{Acknowledgements}
We acknowledge valuable discussions with 
J. L. Cardy, J.-S. Caux, and J. Chalker. M.J.B. would like to thank Hiroshima
University for hospitality and Monbusho and EPSRC for financial
support.  N. T. acknowledges the
support from the Grant-in-Aid for Scientific research No. 11216204 by
Monbusho. A.M.T. is grateful to  M. R. Zirnbauer, 
A. B. Zamolodchikov, A. Mirlin, V. Fateev, C. Pepin and especially to 
V. Kravtsov and I. Lerner for  very fruitful
discussions and constructive criticism of the work.
 
\appendix
\section*{}
\label{sect:super}
\subsection{Supergroups}
In this section we provide a brief introduction to supergroups. For
the reader interested in further details we refer them to the
book \cite{Efetov:chaos} and references therein. For the purposes of this paper we consider matrices of the form
\begin{equation}
M=\begin{pmatrix}a & \sigma\\ \rho & b\end{pmatrix}
\end{equation}
where $a$ and $b$ are respectively $n\times n$ and $m\times m$ matrices containing
even elements of a Grassmann algebra, and $\sigma$ and $\rho$ are respectively $n\times
m$ and $m\times n$ matrices consisting of odd elements of a Grassmann
algebra. Such matrices are called {\em supermatrices}.
The set of such complex (respectively real) square supermatrices is
denoted by M($m|n;\Bbb C$) (respectively M($m|n;\Bbb R$)). The supertrace of M is
defined as 
\begin{equation}
\str(M)=\tr(a)-\tr(b)
\end{equation}
where the symbol $\tr$ stands for the conventional trace. This
definition provides the invariance under cyclic permutations:
\begin{equation}
\str (M_1 M_2\cdots M_n)=\str (M_nM_1\cdots M_{n-1})
\end{equation}
for arbitrary supermatrices $M_1,\cdots, M_n$.
The superdeterminant (or Berezinian) of M is defined as
\begin{equation}
{\rm sdet}(M)=\frac{{\rm det}(a-\sigma b^{-1}\rho)}{{\rm det}(b)}.
\end{equation}
This definition provides the factorization property of the
superdeterminant:
\begin{equation}
{\rm sdet}(M_1M_2\cdots M_n)={\rm sdet}(M_1){\rm sdet}(M_2)\cdots{\rm sdet}(M_n).
\end{equation} 

The {\em general linear supergroup} GL($m|n;\Bbb C$)
(respectively GL($m|n;\Bbb R$)) is
the supergroup of invertible complex (respectively real) supermatrices, the group
composition law being the product of supermatrices. In particular, an arbitrary
element $g\in GL(n|n)$ admits the {\em Gauss decomposition}\cite{Isidro:gltop}
\begin{equation}
\label{Gauss}
g=\begin{pmatrix}{\openone} & 0 \\ \lambda & {\openone}\end{pmatrix}\begin{pmatrix}A &0 \\ 0 & B\end{pmatrix}\begin{pmatrix}{\openone} & \chi \\ 0 & {\openone}\end{pmatrix}
\end{equation}
where $A$ and $B$ ($\lambda$ and $\chi$) are arbitrary $n\times n$
Grassmann-even (respectively Grassmann-odd) matrices. 

The {\em special linear supergroup} SL($m|n;\Bbb C$) is the
subsupergroup of supermatrices $M\in$ GL($m|n;\Bbb C$) such that ${\rm
sdet}(M)=1$. This requirement forces ${\rm det}(A)={\rm
det}(B)$ in the Gauss decomposition (\ref{Gauss}). Thus, an
arbitrary element $g\in$ SL($n|n;\Bbb C$), may be decomposed as
\begin{equation}
\label{slgauss}
g=\exp({\Phi})\begin{pmatrix}{\openone} & 0 \\ \lambda & {\openone}\end{pmatrix}\begin{pmatrix}a &0 \cr 0 & b\end{pmatrix}\begin{pmatrix}{\openone} & \chi \\ 0 & {\openone}\end{pmatrix}
\end{equation} 
where $\Phi\in\Bbb C$, and $a$ and $b$ are arbitrary $n\times n$ {\em unimodular}
matrices (${\rm det }(a)={\rm det }(b)=1$) with Grassmann-even
entries.
\subsection{Superalgebras}
In addition to supergroups, one may also introduce the notion of a Lie
superalgebra, in which the Lie bracket is replaced by a generalised bracket
 (commutator/anticommutator) that depends on whether the generators
considered are `bosonic' or `fermionic'. A particularly useful dictionary of Lie superalgebras
has been compiled in \cite{Frappat:dict}. 

The superalgebra gl($2|2$) is generated by sixteen independent
matrices, which in a suitable basis may be chosen as the twelve
off-diagonal matrices consisting of a single entry of unity, 
together with the matrices 
\begin{equation}
{T}_{{\rm 13}}=\begin{pmatrix}{\bf\sigma}^{3} & 0 \\ 0 &
0\end{pmatrix}\quad {T}_{{\rm 14}}=\begin{pmatrix}0 & 0 \\ 0 &
{\bf\sigma}^{3}\end{pmatrix}\quad {I}=\begin{pmatrix}{\openone}& 0 \\ 0 & {\openone}\end{pmatrix}\quad {\Sigma}=\begin{pmatrix}{\openone}& 0 \\ 0 & -{\openone}\end{pmatrix}
\end{equation}
where $\sigma^3={\rm diag}(1,-1)$, and ${\openone}$
denotes the $2\times2$ unit matrix.
 The Lie algebra  sl($2|2$) is generated by supertraceless matrices,
and may be obtained from the above representation by 
removing ${\Sigma}$. An arbitrary
sl($2|2$) valued current may be expanded as
\begin{equation}
{J}(z)=i(z){I}+\sigma(z)\Sigma +\sum_{i=1}^{14}t_{i}(z){T}_i.
\end{equation}
The Lie algebra sl($2|2$) contains a non-trivial centre; the unit
matrix $I$ commutes with all the generators. The algebra psl(2$|$2) is
obtained by removing this generator. 
One may isolate the contribution to the sl($2|2$) current arising  from the
identity component by the following relation
\begin{equation}
\label{isolate}
i(z)=\frac{1}{4}\str({\Sigma}{J})=\frac{1}{4}\tr{J}
\end{equation}
\subsection{Scaling Dimensions}
\label{multi}
We take a $d$-dimensional sample of material of side $L$ and divide it
into $N=(L/l)^d$ boxes of side $l$. The probability of finding an
electron in box $i$ (the so called {\em box-probability}) is given by
\begin{equation}
P_i=\int d^d{\bf r}_i\,|\psi({\bf r})|^2.
\end{equation}
The average value of the box probability scales in the following manner
\begin{equation}
\label{aveprob}
\langle P \rangle =\frac{1}{N}\sum_{i=1}^{N}P_i\sim \left(\frac{l}{L}\right)^d
\end{equation}
for all normalized wavefunctions ($\sum_i^NP_i=1$) and is not useful
in distinguishing between localized, extended, and critical
wavefunctions. We are led to consider the scaling of the {\em moments}
of the box-probabilities
\begin{equation}
\label{scaling}
\langle P^q\rangle\sim\left(\frac{l}{L}\right)^{d+\tau_q}
\end{equation}
which serves to define new exponents $\tau_q$. Consistency with
(\ref{aveprob}) requires\footnote{
For a uniform electron distribution $\langle P^q \rangle_{\rm uniform}
=\frac{1}{N}\sum_{i=1}^{N}\left(\frac{1}{N}\right)^q\sim
\left(\frac{l}{L}\right)^{dq}$ and one finds that $\tau_q=(q-1)d$. It is conventional to define {\em
generalised dimensions}, $D_q$, such that $\tau_q\equiv(q-1)D_q$. The scaling relation (\ref{scaling}) may equivalently be written $\langle P^q\rangle\sim\left(\frac{l}{L}\right)^{d+(q-1)D_q}$. We note that (\ref{valtau}) implies $D_0=d.$}
\begin{equation}
\label{valtau}
\tau_1=0 \quad \tau_0 = -d.
\end{equation}
It is often advantageous (especially in numerical simulations) to introduce the Legendre transform, $f$, of
$\tau(q)$,
\begin{equation}
\label{legendre}
f(\alpha_q)\equiv q\alpha_q-\tau_q,
\end{equation}
where
\begin{equation}
\label{alpha}
        \alpha_q\equiv\frac{d\tau_q}{dq}.
\end{equation}
It follows that
\begin{equation}
\frac{df(\alpha)}{d\alpha}=q
\end{equation}
Thus, the function $f(\alpha)$ has a maximum at $q=0$. Using
(\ref{valtau}) and (\ref{legendre}),
\begin{equation}
\label{max}
f(\alpha)_{\rm max}=f(\alpha_0)=d.
\end{equation}
The slope of $f(\alpha)$ is unity at the value of $\alpha$
corresponding to $q=1$,
\begin{equation}
\label{slope}
\left.\frac{df(\alpha)}{d\alpha}\right|_{q=1}=1.
\end{equation}
Combining (\ref{valtau}) and (\ref{legendre}) we also find that,
\begin{equation}
\label{value}
f(\alpha_1)=\alpha_1
\end{equation}
The constraints (\ref{max}), (\ref{slope}) and (\ref{value}) allow us
to write a parabolic {\em approximation} in terms of one parameter,
$\alpha_0$ (the position of the maximum)\footnote{Assume
$f(\alpha)=d-A(\alpha-\alpha_0)^2$ where $A$ is a constant to be
determined; this takes into account (\ref{max}). Condition
(\ref{slope}) implies $\alpha_1=\alpha_0-1/(2A)$. Substituting this into
the equation following from (\ref{value}), namely
$\alpha_1=d-A(\alpha_1-\alpha_0)^2$, determines $A$.}
\begin{equation}
f(\alpha)=d-\frac{(\alpha-\alpha_0)^2}{4(\alpha_0-d)}.
\end{equation}
This parabolic approximation together with the definitions
(\ref{legendre}) and (\ref{alpha})
gives rise to the relation 
\begin{equation}
q\frac{d\tau}{dq}-\tau=d-\frac{(\frac{d\tau}{dq}-\alpha_0)^2}{4(\alpha_0-d)}
\end{equation}
This equation is solved {\em exactly} by the polynomial 
\begin{equation}
\tau_q=-d+\alpha_0q-(\alpha_0-d)q^2.
\end{equation}
Recalling the scaling relation (\ref{scaling}) one obtains
\begin{equation}
\langle P^q\rangle\sim\left(\frac{l}{L}\right)^{\alpha_0q-(\alpha_0-d)q^2}.
\end{equation}
The local density of states is given by
\begin{equation}
\rho=\delta^{-1}P\sim L^dP.
\end{equation}
Hence
\begin{equation}
\langle \rho^q\rangle\sim\frac{1}{L^{(\alpha_0-d)q(1-q)}}.
\end{equation}
Numerical simulations for $d=2$ give $\alpha_0\approx 2.28$.
\subsection{Representations of the SL(2,${\mathbf {\Bbb R}}$) Group}
\label{SLrep}

The Lie algebra sl$(2)$ is generated by three independent traceless matrices,
which, in the {\em spin basis} may be taken as
\begin{eqnarray}
J^-=\left(
\begin{array}{cc}
0 & 0 \\ 1 & 0
\end{array}\right),\quad J^0=\frac{1}{2}\left(\begin{array}{cc}1 & 0
\\ 0 & -1
\end{array}\right),\quad J^+=
\left(\begin{array}{cc}0 & 1 \\ 0 & 0
\end{array}\right).
\end{eqnarray}
Their commutation relations read
\begin{equation}
\label{spincom}
[J^+,J^-]=2J^0,\quad [J^0,J^\pm]=\pm J^\pm,
\end{equation}
and should be familiar from the theory of angular momentum. It is
readily seen that the differential operators
\begin{equation}
J^-=\frac{\partial}{\partial y}, \quad J^0=y\frac{\partial}{\partial
y}-j, \quad J^+=2jy-y^2\frac{\partial}{\partial y}
\end{equation}
obey the same commutation relations (\ref{spincom}) when acting on the
space of differentiable functions.

In this differential realisation, the monomial $y^{j+m}$ plays the
r{\^o}le of the state vector $|j,m\rangle$. One observes that
$\partial/\partial y$ acts as a kind of lowering operator, and
$y$ acts as a kind of raising operator.
It is convenient to modify our basis slightly by defining $t^{+1}=-J^+$,
$t^0=J^0$, $t^{-1}=J^{-}$. This enables us to write the generators in
the following compact form
\begin{equation}
\label{sl2rgen}
{t^l=y^{(l+1)}\frac{\partial}{\partial y}-(l+1)jy^l; \quad l=-1,0,+1}
\end{equation}
Their commutation relations read
\begin{equation}
[t^+,t^-]=-2t^0, \quad [t^0,t^{\pm}]=\pm t^\pm
\end{equation}
That is to say we have structure constants,
\begin{equation}
{f^{0+}}_+=1=-{f^{0-}}_-,\quad {f^{+-}}_0=-2
\end{equation}
The Killing form reads (with the ordering $+,0,-$, and $c_{\rm v}=2$) 
\begin{eqnarray}
\eta^{ab}=\left(\begin{array}{ccc} 0 & 0 & 2 \\ 0 & -1 & 0 \\ 2 &
0 & 0 \end{array}\right), \quad \eta_{ab}=\left(\begin{array}{ccc} 0 & 0 & \frac{1}{2} \\ 0 & -1 & 0 \\ \frac{1}{2} &
0 & 0 \end{array}\right)
\end{eqnarray}
More generally, one can construct representations 
 defining their action  on functions 
$f(y)$:
\begin{equation}
\hat A f(y) = |cy +d|^{2j}\,{\rm sign}\,(cy + d)^{2\epsilon}f\left(\frac{ay + b}{cy +
d}\right)
\end{equation}
where $ad - cb = 1$. 

The operator $\rho^q$ discussed in the text 
belongs to the $j = - q$ representation  and
is  annihilated by $t^-$:
\begin{equation}
t^+_{j = -q}\rho^q = 0, ~~ \rho^qt^- = 0
\end{equation}
In other words one may perform the following expansion
\begin{equation}
\rho^q(y,\bar y) = \sum_{k,\bar k = 0}y^{- 2q - k}{\bar y}^{\bar
k}\Phi^{(-q)}_{k,\bar k}
\end{equation}
In this representation $t^3$ is diagonal :
\begin{equation}
t^3\Phi_{k,\bar k} = - (q + k)\Phi_{k,\bar k}  
\end{equation}
The operator $\Phi_0$ belongs to the representation where $t^3$ cannot
be diagonalized:
\begin{equation}
t^+\Phi_0 = \mu\Phi_0, ~~ \Phi_0 t^- = \mu\Phi_0
\end{equation}
For more information see, for example \cite{Carmeli}.

\subsection{The Knizhnik-Zamolodchikov Equation}
The stress-energy tensor of WZNW models is proportional to the scalar
product of currents. Therefore   the Virasoro generator with  $n=-1$
acting on any primary field gives 
\begin{equation}
L_{-1}|\phi_i\rangle=\frac{2}{k+c_{\rm
v}}\eta_{ab}J^a_{-1}J^b_0|\phi_i\rangle=\frac{-2}{k+c_{\rm v}}\eta_{ab}J^a_{-1}t^b_i|\phi_i\rangle
\end{equation}
We consider the insertion of the zero vector
\begin{equation}
|\chi\rangle=\left[L_{-1}+\frac{2}{k+c_{\rm v}}\eta_{ab}J^a_{-1}t^b_i\right]|\phi_i\rangle=0
\end{equation}
inside the correlation function of a set of primary fields. We note
that the insertion of the operator $J_1^a$ in the correlator can be
expressed as
\begin{eqnarray}
\langle\phi_1(z_1)\cdots (J_{-1}^a\phi_i)(z_i)\cdots\phi_n(z_n)\rangle&=&\frac{1}{2\pi i}\oint_{z_i}\frac{dz}{z-z_i}\langle 
J^a(z)\phi_1(z_1)\cdots\phi_n(z_n)\rangle \\
& =& -\sum_{j\neq i}\frac{1}{2\pi i}\oint_{z_i}\frac{dz}{(z-z_i)}\frac{t_j^a}{(z-z_j)}\langle \phi_1(z_1)\cdots\phi_n(z_n)\rangle\\
& = & -\sum_{j\neq i}\frac{t_j^a}{(z_i-z_j)}\langle \phi_1(z_1)\cdots\phi_n(z_n)\rangle
\end{eqnarray}
Therefore
\begin{eqnarray}
\langle\phi_1(z_1)\cdots \chi(z_i)\cdots\phi_n(z_n)\rangle & = &\left[\partial_{z_i}-\frac{2}{k+c_{\rm v}}\sum_{j\neq
i}\frac{\eta_{ab}\,t^a_i\otimes t^b_j}{z_i-z_j}\right]\langle \phi_1(z_1)\cdots\phi_n(z_n)\rangle
\end{eqnarray}
and by construction this must vanish:
\begin{equation}
\label{kz}
{
\left[\partial_{z_i}-\frac{2}{k+c_{\rm v}}\sum_{j\neq
i}\frac{\eta_{ab}\,t^a_i\otimes t^b_j}{z_i-z_j}\right]\langle \phi_1(z_1)\cdots\phi_n(z_n)\rangle=0}.
\end{equation}
This is the {\em Knizhnik--Zamolodchikov equation}. The solutions to
this equation are the correlation functions of primary fields.

\subsection{BRST Invariance and {\bf $\Delta=0$} Conformally Invariant
Deformation}
\label{BRST} 
Let us consider a WZNW model 
 at 
level $-k$\footnote{The notation is choosen in such a way that for $k
> 0$ the action for  the SL($N,{\Bbb C}$)/SU($N$) model is positive.}
deformed by its kinetic term:
\begin{equation}
S(g)=S_{WZNW}(g^{-1})~-~\epsilon\int d^2 z\Omega(z,\bar z)
\label{expansion}\end{equation}
where 
\begin{equation}
\Omega(z,\bar z)=J^a(z)\bar J^b(\bar z)\phi^{ab}(z,\bar z),\label{Omega}\end{equation}
and
\begin{equation}
J= {1\over2}kg^{-1}\partial g,~~~~\bar J= {1\over2}k\bar\partial gg^{-1},
~~~~\phi^{ab}=\mbox{tr}g^{-1}t^agt^b.
\end{equation}

The perturbation operator $\Omega$ possesses 
a number of interesting properties. 
Let us compute the commutator of the Ka{\v c}--Moody current $J^a(z)$
with the operator $\Omega$. Denoting its chiral part $O(z)$ we get 
\begin{eqnarray}
[J^a(y),O(z)]&=&\oint {d\zeta\over2\pi i}{1\over\zeta -z}\{[J^a(y),J^b(\zeta)]\phi^b(z)
+J^b(\zeta)[J^a(y),\phi^b(z)]\}\nonumber\\
&=&\oint {d\zeta\over2\pi i}{1\over\zeta -z}\{f^{ab}_cJ^c(\zeta)\phi^b(z)\delta(y,\zeta)+
{1\over2}k\phi^a(z)\delta'(y,\zeta)+f^{ab}_cJ^b(\zeta)\phi^c(z)\delta(y,z)\}\nonumber\\
&=&\oint {d\zeta\over2\pi i}{1\over\zeta -z}\left\{{f^{ab}_cf^{cb}_d\over\zeta - 
z}\phi^d(z)\delta(y,\zeta)+{1\over2}k\phi^a(z)\delta'(y,\zeta)+{f^{ab}_cf^{bc}_d\over\zeta -
z}\phi^d(z)\delta(y,z)\right\}\nonumber\\
&+&\oint {d\zeta\over2\pi i}{1\over\zeta -z}[f^{ab}_c\Psi^{cd}(z)\delta(y,\zeta)+
f^{ac}_b\Psi^{cb}(z)\delta(y,z)].
\end{eqnarray}
Here
\begin{equation}
\Psi^{cb}(z)=:J^c(z)\phi^b(z):.
\end{equation}
By taking contour integrals, we obtain
\begin{equation}
[J^a(y),O(z)]=(-{1\over2}k+c_V)\phi^a(z)\delta'(y,z).
\end{equation}
 Thus for the entire operator $\Omega$ we get 
find
\begin{equation}
[J^a(z),\Omega(w,\bar w)]=\left(-{1\over2}k~+~c_V\right)\bar J^b(\bar w)\phi^{ab}(w,\bar w)\delta'(z,w).
\label{commutator}\end{equation}
For the SL($N,{\Bbb C}$)/SU($N$) model where $c_V = N$ 
 this commutator vanishes when $k=2N.\label{k}$ (as in the main text the level 
$k$ is defined  such  that the central
charge is given by $C = k(N^2 - 1)/(k - N)$). 

In other words, when $k=2N$, the operator $\Omega$ is invariant under
the Ka{\v c}--Moody symmetry. 
Since the Virasoro operators are quadratic combinations of the Ka{\v c}--Moody currents, the operator 
$\Omega$ is automatically invariant under the conformal transformations. In particular,
\begin{equation}
[L_0,\Omega]=\Delta\Omega=0.\label{L_0}\end{equation}
Indeed,
\begin{equation}
\Delta(k=2N)=0.\label{Delta}\end{equation}
The operator $\Omega$ forms a closed OPE algebra
\begin{equation}
\Omega(z)\Omega(0)=\Omega(0).\label{OPE}\end{equation}
The constant term on the right hand side is absent because the
operator $\Omega$ has zero norm; this follows from equation (\ref{commutator}).

Thus, a
the SL($N,{\Bbb C}$)/SU($N$) WZNW model at 
level $k=2N$ perturbed by the operator $\Omega$ must be conformally
invariant for arbitrary parameter $\epsilon$. 
This can be seen in the following way: 
The property (\ref{OPE}) implies that away from the conformal point the following relation holds
\begin{equation}
\bar\partial T=\partial\Theta,\end{equation}
where $T$ and $\Theta$ are the components of the stress-energy
tensor. 
Moreover, 
the trace is given as follows
\begin{equation}
\Theta=\beta_\Omega~\Omega,\label{Theta}\end{equation}
where $\beta_\Omega$ is the renormalization group beta-function of the coupling $\epsilon$. 
Since $\Omega$ is not a marginal operator, $\beta_\Omega\ne0$. However, the following is true
\begin{equation}
\partial\Theta=[L_{-1},\Theta]=\beta_\Omega[L_{-1},\Omega].\label{derivative}\end{equation}
Since $[L_n,\Omega]=0 \label{allzero}$
for any $n$, we arrive at
\begin{equation}
\bar\partial T=0,\label{holomorphic}\end{equation}
i.e. the stress-energy tensor component $T$ is still a holomorphic function even away from the critical 
point. This means that the perturbed theory remains conformal for an
arbitrary perturbation. This also means that the most stable point in
this case is not $k = 1$ as for the SU(N) WZNW model, but $k = 2N$. 

 The described  possibility is not realized for the PSL(2$|$2) model 
where $c_V = 0$. Though the scaling dimension of the $\Omega$ operator in this
case is equal to 2 and it may appear being  exactly marginal, the fact
that it does not commute with $J$'s may be  an indication that 
the  line of critical points obtained 
 by studying  {\it perturbative} 
series \cite{Berkovits:ramramflux,Bershadsky:psl} is destroyed by
 non-perturbative effects. This is an open interesting question.


\begin{thebibliography}{10}

\bibitem{VonKliz:iqhe}
K.~von Klitzing, G.~Dorda, and M.~Pepper,
\newblock Phys. Rev. Lett. {\bf 45}, 494 (1980).

\bibitem{Tsui:iqhe}
D.~C. Tsui, H.~L. St{\"o}rmer, and A.~C. Gossard,
\newblock Phys. Rev. Lett. {\bf 48}, 1559 (1982).

\bibitem{Fransesco:cft}
P.~D. Francesco, P.~Mathieu, and D.~S{\'e}n{\'e}chal,
\newblock {\em Conformal Field Theory} (Springer, 1997).

\bibitem{Ketov:cft}
S.~V. Ketov,
\newblock {\em Conformal Field Theory} (World Scientific, 1995).

\bibitem{Tsvelik:boson}
A.~O. Gogolin, A.~A. Nersesyan, and A.~M. Tsvelik,
\newblock {\em Bosonization in Strongly Correlated Systems} (Cambridge
  University Press, 1999).

\bibitem{Levine:sigmod}
H.~Levine, S.~B. Libby, and A.~M.~M. Pruisken,
\newblock Phys. Rev. Lett. {\bf 51}, 1915 (1983).

\bibitem{Pruisken:sigmod}
A.~M.~M. Pruisken,
\newblock Nucl. Phys. {\bf B235}, 277 (1984).

\bibitem{Khmel:twopar}
D.~E. Khmelnitskii,
\newblock JETP. Lett. {\bf 38}, 552 (1983).


\bibitem{Pruisken3}
H.~Levine, S.~B. Libby, and A.~M.~M. Pruisken,
\newblock Nucl. Phys.  (1984).

\bibitem{Knizmoz}
V.~G. Knizhnik and A.~Y. Morozov,
\newblock JETP. Lett. {\bf 39}, 241 (1984).

\bibitem{Efetov:super}
K.~B. Efetov,
\newblock Adv. Phys. {\bf 32} (1983).

\bibitem{Efetov:chaos}
K.~Efetov,
\newblock {\em Supersymmetry in Disorder and Chaos} (Cambridge University
  Press, 1997).

\bibitem{Weiden:supersig}
H.~A. Weidenmuller,
\newblock Nucl. Phys. {\bf B290}, 87 (1987).


\bibitem{Seiberg:seiberg}
N.~Seiberg,
\newblock Nucl. Phys. {\bf B435}, 129 (1995).

\bibitem{Zamol}
A.~B. Zamolodchikov,
\newblock JETP. Lett. {\bf 43}, 565 (1986).

\bibitem{Ludwig}
A.~A. Ludwig and J.~L. Cardy,
\newblock Nucl. Phys. {\bf 285}, 687 (1987).

\bibitem{Chalker:net}
J.~T. Chalker and P.~D. Coddington,
\newblock J. Phys. {\bf C21}, 2665 (1987).



\bibitem{DhLee:94}
D.-H. Lee,
\newblock Phys. Rev. {\bf 50}, 788 (1994).

\bibitem{KondMar:97}
N.~Read,
\newblock Unpublished.


\bibitem{Read:super}
J.~Kondev and J.~B. Marston,
\newblock Nucl. Phys. {\bf B497}, 639 (1997).


\bibitem{Zirn:super}
M.~R. Zirnbauer,
\newblock Annalen der Physik {\bf 3}, 513 (1994).

\bibitem{Zirn:97}
M.~R. Zirnbauer,
\newblock J. Math. Phys. {\bf 38}, 2007 (1997).

\bibitem{Zirn:99}
M.~R. Zirnbauer,
\newblock J. Math. Phys. {\bf 40}, 2197 (1999).




\bibitem{Zirnbauer:integer}
M.~R. Zirnbauer,
\newblock hep-th/9905054.

\bibitem{guruswamy} S. Guruswamy. A. LeClair and A. W. W. Ludwig,
cond-mat/9909143.

\bibitem{Zirnbauer86} M.~R. Zirnbauer, Nucl. Phys. B{\bf 265}, 375
(1986). 
 
\bibitem{Mirlin:onedim}
A.~D. Mirlin,
\newblock J. Math. Phys {\bf 38}, 1888 (1997).


\bibitem{Falko}
V.~I. Fal'ko and K.~B. Efetov,
\newblock Phys. Rev. {\bf 17}, 413 (1995).



\bibitem{Wegner}
F.~Wegner,
\newblock {\em Localization and Metal-Insulator Transition} (Plenum, N.Y.,
  1985), .

\bibitem{Janssen:mulfrac}
M.~Janssen, M.~Metzler, and M.~R. Zirnbauer,
\newblock Phys. Rev. {\bf B59}, 836 (1999).

\bibitem{Caux:disferm}
J.-S. Caux, N.~Taniguchi, and A.~M. Tsvelik, 
\newblock Phys. Rev. Lett. {\bf 80}, 1276 (1998); 
Nucl. Phys. {\bf B525}, 621 (1998). J.-S. Caux, Phys. Rev. Lett. {\bf
81}, 4196 (1998). 

\bibitem{MF} Y.~V. Fyodorov and A.~D. Mirlin, Int. J. Phys. B{\bf 8},
3795 (1994). 

\bibitem{Carmeli}
M.~Carmeli,
\newblock {\em Group Theory and General Relativity} (McGraw-Hill,
1977).

\bibitem{Berkovits:ramramflux}
N.~Berkovits, C.~Vafa, and E.~Witten,
\newblock hep-th/9902098.

\bibitem{Bershadsky:psl}
M.~Bershadsky, S.~Zhukov, and A.~Vaintrob,
\newblock hep-th/9902180 v2.

\bibitem{Goddard:coset}
P.~Goddard, A.~Kent, and D.~Olive,
\newblock Phys. Lett. {\bf 152B}, 88 (1985).

\bibitem{Kogan:gauged}
I.~I. Kogan, A.~Lewis, and O.~A. Soloviev,
\newblock Int. J. Mod. Phys. {\bf A13}, 1345 (1998)
\bibitem{Poly:multi}
A.~M. Polyakov and P.~B. Wiegmann,
\newblock Phys. Lett. {\bf 141B}, 223 (1984).

\bibitem{Geras:free}
A.~Gerasimov, A.~Morozov, M.~Olshanetsky, A.~Marshakov, and S.~Shatashvili,
\newblock Int. J. Mod. Phys. {\bf A5}, 2495 (1990).

\bibitem{Isidro:gltop}
J.~M. Isidro and A.~V. Ramallo, 
\newblock Nucl. Phys. {\bf B414}, 715 (1994).

\bibitem{Poly:nonab}
A.~M. Polyakov and P.~B. Wiegmann,
\newblock Phys. Lett. {\bf 131B}, 121 (1983).

\bibitem{FZ}
V.~Fateev and A.~B. Zamolodchikov,
\newblock Sov. Nucl. Phys. {\bf 43}, 1031  (1986).

\bibitem{KT}
I.~I. Kogan and A.~M. Tsvelik, hep-th/9912143

\bibitem{Frappat:dict}
L.~Frappat, P.~Sorba, and A.~Sciarrino,
\newblock hep-th/9607161.

\end{thebibliography}

\end{document}